\documentclass[aps,prd,twocolumn,superscriptaddress, nofootinbib,floats,floatfix]{revtex4-1}
\usepackage{bm}
\usepackage{amsmath}
\usepackage{amsfonts}
\usepackage{aas_macros}
\usepackage{graphicx}
\usepackage[hidelinks]{hyperref}
\usepackage{booktabs}
\usepackage{color}
\usepackage[dvipsnames]{xcolor}
\usepackage[capitalise]{cleveref}
\usepackage[caption=false]{subfig}
\usepackage{paralist}
\usepackage{acro}

\newcommand\sun{\hbox{$\odot$}}

\newcommand\csiborg{\texttt{CSiBORG} }
\newcommand\borg{\texttt{BORG} }
\newcommand\nside{\texttt{nside}}

\newcommand\healpix{\textsc{HEALPix} }
\newcommand\healpy{\textsc{healpy} }
\newcommand\emcee{\textsc{emcee} }
\newcommand\clumpy{\textsc{clumpy} }
\newcommand\fermipy{\textsc{FermiPy} }

\newcommand{\dd}{\mathrm{d}}
\newcommand\numberthis{\addtocounter{equation}{1}\tag{\theequation}}

\creflabelformat{equation}{#2\textup{#1}#3}

\newcommand{\Mpch}{\ensuremath{h^{-1}\;\text{Mpc}}}
\DeclareMathOperator{\erf}{erf}
\newcommand{\sigv}{\ensuremath{\left< \sigma v \right>}}
\newcommand{\Msun}{\ensuremath{\mathrm{M}_\odot}}

\DeclareAcronym{DM}{short = DM, long  = dark matter}
\DeclareAcronym{CMB}{short = CMB, long  = Cosmic Microwave Background}
\DeclareAcronym{ICs}{short = ICs, long  = initial conditions}
\DeclareAcronym{BBN}{short = BBN, long  = Big Bang Nucleosynthesis}
\DeclareAcronym{MCMC}{short = MCMC, long = Markov Chain Monte Carlo}
\DeclareAcronym{SPH}{short = SPH, long = smooth particle hydrodynamics}
\DeclareAcronym{CIC}{short = CIC, long = cloud-in-cell}

\begin{document}

\title{Constraints on dark matter annihilation and decay from the large-scale structure of the nearby universe}

\author{D. J. Bartlett}
\email{deaglan.bartlett@physics.ox.ac.uk}
\affiliation{Astrophysics, University of Oxford, Denys Wilkinson Building, Keble Road, Oxford, OX1 3RH, UK}\affiliation{CNRS \& Sorbonne Universit\'{e}, Institut d’Astrophysique de Paris (IAP), UMR 7095, 98 bis bd Arago, F-75014 Paris, France}
\author{A. Kosti\'{c}}
\email{akostic@mpa-garching.mpg.de}
\affiliation{Max Planck Institute for Astrophysics, Karl-Schwarzschild-Stra{\ss}e 1, 85748 Garching, Germany}
\author{H. Desmond}
\affiliation{Astrophysics, University of Oxford, Denys Wilkinson Building, Keble Road, Oxford, OX1 3RH, UK}
\affiliation{McWilliams Center for Cosmology, Department of Physics, Carnegie Mellon University, 5000 Forbes Ave, Pittsburgh, PA 15213}
\affiliation{Institute of Cosmology \& Gravitation, University of Portsmouth, Dennis Sciama Building, Portsmouth, PO1 3FX, UK}
\author{J. Jasche}
\affiliation{The Oskar Klein Centre, Department of Physics, Stockholm University, AlbaNova University Centre, SE 106 91 Stockholm, Sweden}\affiliation{CNRS \& Sorbonne Universit\'{e}, Institut d’Astrophysique de Paris (IAP), UMR 7095, 98 bis bd Arago, F-75014 Paris, France}
\author{G. Lavaux}
\affiliation{CNRS \& Sorbonne Universit\'{e}, Institut d’Astrophysique de Paris (IAP), UMR 7095, 98 bis bd Arago, F-75014 Paris, France}

\begin{abstract}
    Decaying or annihilating dark matter particles could be detected through gamma-ray emission from the species they decay or annihilate into. This is usually done by modelling the flux from specific dark matter-rich objects such as the Milky Way halo, Local Group dwarfs, and nearby groups. However, these objects are expected to have significant emission from baryonic processes as well, and the analyses discard gamma-ray data over most of the sky. Here we construct full-sky templates for gamma-ray flux from the large-scale structure within $\sim$200 Mpc by means of a suite of constrained $N$-body simulations (\textsc{CSiBORG}) produced using the Bayesian Origin Reconstruction from Galaxies algorithm. Marginalising over uncertainties in this reconstruction, small-scale structure, and parameters describing astrophysical contributions to the observed gamma-ray sky, we compare to observations from the \textit{Fermi} Large Area Telescope to constrain dark matter annihilation cross sections and decay rates through a Markov Chain Monte Carlo analysis. We rule out the thermal relic cross section for $s$-wave annihilation for all $m_\chi \lesssim 7 {\rm \, GeV}/c^2$ at 95\% confidence if the annihilation produces gluons or quarks less massive than the bottom quark. We infer a contribution to the gamma-ray sky with the same spatial distribution as dark matter decay at $3.3\sigma$. Although this could be due to dark matter decay via these channels with a decay rate $\Gamma \approx 6 \times 10^{-28} {\rm \, s^{-1}}$, we find that a power-law spectrum of index $p=-2.75^{+0.71}_{-0.46}$, likely of baryonic origin, is preferred by the data.
\end{abstract}

\maketitle

\section{Introduction}

The actual particle nature of \ac{DM} is not yet known, despite it having 5 times the average density of baryonic matter. Theoretically favourable candidates are weakly interacting massive particles (WIMPs) \citep{Steigman_1985,Kamionkowski_1998,Bertone_2005}, including, but not limited to, the lightest supersymmetric particle in supersymmetric theories. If these WIMPs are thermal relics of the early Universe, then the current abundance of \ac{DM} suggests that the self-annihilation cross section should be \hbox{$\sigv_{\rm th} \approx 3 \times 10^{-26} {\rm \, cm^3 s^{-1}}$} \citep{Steigman_2012} (the thermal relic cross section), which is suspiciously similar to what one would expect for a particle of mass $\sim 0.1-1 {\rm \, TeV}$ with a coupling comparable to the electroweak coupling of the Standard Model (SM). Recent experiments find an anomalous muon magnetic moment, as measured by the ``Muon $g-2$'' experiment \citep{Muon_2021}, and a mass for the $W$ boson that is higher than expected \citep{CDF_2022}. This
further motivates probing physics beyond the SM and thus the search for \ac{DM} candidates. Detection of such particles could therefore solve some of the fundamental questions of particle physics and cosmology.

\ac{DM} annihilation or decay for particles at these masses could be detected through the emission of gamma rays by their decay products. Despite the theoretically low interaction rates - one would expect only a few in $10^{15}$ particles to annihilate per Hubble time in the present Universe \citep{Slatyer_LesHouches} - the vast quantities of \ac{DM} on cosmological scales makes these processes potentially detectable in the state-of-the-art gamma-ray measurements from the \textit{Fermi} Large Area Telescope (\textit{Fermi}-LAT) \citep{Ackermann_2010}. Indeed, the excess of observed gamma rays toward the galactic centre (Galactic Centre Excess; GCE) \citep{Goodenough_2009,Ajello_2016,Linden_2016} can be fitted well by the annihilation of \ac{DM} \citep{Hooper_2011,Hooper_2011PRD,Hooper_2013PDU,Zhou_2015,Daylan_2016,Cholis_2021,Grand_2022}. There is debate over whether other explanations could also explain the emission. Several studies \citep{Abazajian_2011,OLeary_2015,Petrovic_2015,Lee_2016,Buschmann_2020,Gautam_2021} argue that the GCE can be explained by a population of unresolved point sources, although this is contested \citep{Hooper_2013,Cholis_2015_pulsars,Leane_2019,Cholis_2021}, and some groups \citep{Abazajian_2012,Gordon_2013,Abazajian_2014,Calore_2015} find both \ac{DM} annihilation and other models can fit the data. A combination of these two processes is, of course, possible, and is plausible given the spatial variation of the GCE \citep{Horiuchi_2016}. Other astrophysical processes at the centre of the galaxy \citep{Petrovic_2014,Carlson_2014,Cholis_2015_rays,Gaggero_2015,Macias_2018,Bartels_2018} could also be responsible.
 
Given these conflicting explanations of the GCE, in order to unambiguously detect or rule out \ac{DM} annihilation or decay models, one should determine if an excess of gamma rays is detected from other sources or across the full sky. Previous studies have placed constraints on $\sigv$ through cross-correlation between \textit{Fermi}-LAT data and galaxy \cite[e.g.][]{Hashimoto_2021,Hashimoto_2022} or lensing \citep{Ammazzalorso_2020} catalogues, or by studying nearby dwarf galaxies \citep{dSph_0, dSph, dIrr} or groups \citep{groups}. Moreover, such emission should increase the kinetic energy of baryons, so by considering the impact on the \ac{CMB} \citep{Planck_Slatyer} or galaxy formation \citep{dwarf_heating}, one can rule out  a velocity-independent cross section for thermal relic \ac{DM} particles less massive than $\sim 30{\rm \, GeV}$.

Instead of focusing on a few nearby or massive objects, the aim of this work is to search for the signature of \ac{DM} decay and annihilation across the full sky by forward-modelling the observed gamma-ray sky, as first suggested in \citep{Cuesta_2010}. As proposed in \citep{Ando:2005xg}, anisotropies in the cosmic gamma-ray background could be a signature of \ac{DM} annihilation or decay. This has previously been studied through the two-point correlation function \cite[e.g.][]{Fermi-LAT:2012pez}, which is calibrated with \textit{unconstrained} $N$-body simulations (e.g. \citep{Fornasa:2016ohl} use the  Millennium-II and Aquarius simulations \citep{Boylan-Kolchin_2009,Navarro_2010,Springel_2008}). Instead, we utilise the \csiborg suite of \textit{constrained} $N$-body simulations \citep{Bartlett_2021_VS,antihalos,Hutt_2022,max_zenodo}. The \ac{ICs} for these simulations are chosen to produce final three-dimensional \ac{DM} density fields that are consistent with the observed positions of galaxies in the 2M++ galaxy catalogue. The \ac{ICs} are inferred using the \borg (Bayesian Origin Reconstruction from Galaxies) algorithm \citep{BORG_1,BORG_2,BORG_3,BORG_4,Lavaux_2016}, a fully Bayesian forward model. We use the particle positions in the simulations to produce maps of the expected gamma-ray flux from halos down to $4.38 \times 10^{11} {\rm \, M_{\odot}}$ in mass (although we also model smaller substructures), as well as from \ac{DM} not identified as belonging to halos. The halos are assumed to have Navarro-Frenk-White (NFW) profiles \citep{NFW_1997} and we explicitly model unresolved substructure in a probabilistic manner, since the signal from \ac{DM} annihilation is sensitive to the peaks in the density field.

We do not include a contribution from the Milky Way halo or Local Group dwarf galaxies (which are unresolved in our simulations) in our templates so as to produce constraints entirely from large scale structure. These constraints will be free from many of the systematics affecting searches in particular objects and will reveal the amount of constraining power for \ac{DM} annihilation and decay to be found in various parts of the cosmic web. To account for non-DM effects we include templates for emission due to point sources, galactic emission and an isotropic background. We marginalise over the amplitudes of these templates, as well as the realisations in the \csiborg suite which sample the full \borg posterior in \ac{ICs} of the 2M++ volume. We then compare to \textit{Fermi}-LAT observations via a \ac{MCMC} algorithm. A full-sky field-level inference allows us to capture not only the two-point statistics but implicitly all higher orders, too.

In this work we rule out the thermal relic cross section at 95\% confidence for annihilations that produce gluons or quarks less massive than the bottom quark if \ac{DM} has a mass $m_\chi \lesssim 7 {\rm \, GeV}/c^2$. We find a contribution to the gamma-ray sky with the same spatial distribution as expected from \ac{DM} decay (flux proportional to \ac{DM} density) at $3.3\sigma$ confidence, with a decay rate $\Gamma \approx 6 \times 10^{-28} {\rm \, s^{-1}}$ for these channels. However, a power-law spectrum with an index $p=-2.75^{+0.71}_{-0.46}$ provides a better fit to the data, suggesting a non-\ac{DM} origin. In our fiducial analysis, we do not rule out the thermal relic annihilation cross section at any mass for production of top or bottom quarks; we obtain upper bounds that are half the size (i.e. tighter constraints) if we marginalise over the contribution proportional to the \ac{DM} density, but we do not include this contribution in the fiducial analysis. Our constraints on \ac{DM} decay to leptons are approximately an order of magnitude less stringent than decay to quarks.

This paper is structured as follows. We discuss \ac{DM} annihilation and decay models in \cref{sec:theory} and introduce the gamma-ray data that we use to constrain these models in \cref{sec:Observational data}. Our inference and template construction methods are outlined in \cref{sec:Methods}. The results are presented in \cref{sec:Results} and discussed in \cref{sec:Discussion}, including the potential systematic uncertainties and a comparison to the literature. We conclude in \cref{sec:Conclusions}. Equations throughout the paper use units $\hbar= c = 1$.

\section{Theoretical background}
\label{sec:theory}

We start by assuming that \ac{DM} is made of a single particle, $\chi$, of mass $m_\chi$, whose antiparticle is itself (e.g., Majorana fermions). This particle is assumed to be able to both decay
\begin{equation}
    \chi \to A \bar{A},
\end{equation}
and annihilate
\begin{equation}
    \chi\chi \to A \bar{A},
\end{equation}
to a standard model particle, $A$, and its antiparticle, $\bar{A}$. The annihilation of the produced particles would lead to gamma-ray emission at some energy $E_\gamma$, which one could detect. If these processes occur via channel $i$ with branching ratio ${\rm Br}_i$, then the photon flux for annihilation per unit density squared
at redshift $z$ is \citep{Lisanti_2018}
\begin{equation}
    \label{eq:Phi_pp ann}
    \frac{\dd\Phi_{\rm pp}^{\rm ann}}{\dd E_\gamma} = \frac{\sigv}{8 \pi m_\chi^2} \sum_i {\rm Br}_i  \left. \frac{\dd N^{\rm ann}_i}{\dd E_\gamma^\prime} \right\vert_{E_\gamma^\prime = E_\gamma \left( 1 + z \right)},
\end{equation}
and for decay per unit density
\begin{equation}
    \label{eq:Phi_pp dec}
    \frac{\dd \Phi_{\rm pp}^{\rm dec}}{\dd E_\gamma} = \frac{\Gamma}{4 \pi m_\chi} \sum_i {\rm Br}_i  \left. \frac{\dd N^{\rm dec}_i}{\dd E_\gamma^\prime} \right\vert_{E_\gamma^\prime = E_\gamma \left( 1 + z \right)},
\end{equation}
where $\sigv$ is the thermally averaged cross section, $\tau = 1 / \Gamma$ is the lifetime of the particle, and $\dd N_i / \dd E_\gamma$ is the photon energy distribution for channel $i$. Throughout this work we assume $s$-wave annihilation so the parameter $\sigv$ is assumed to be a constant, i.e. independent of $v$. If $\chi$ is not its own antiparticle (e.g., Dirac fermions), the annihilation flux is half of this value, provided there is no matter-antimatter asymmetry. Since we do not \textit{a priori} know which branching ratios to use, in this work we assume that the annihilation or decay occurs via a single channel; however, our analysis can be trivially extended to multiple channels.

Since these results apply at unit density, we must now take into account the integrated \ac{DM} density along the line of sight. By introducing the $J$ factor
\begin{equation}
    \label{eq:J def}
    \frac{\dd J}{\dd \Omega} = \int  \rho_{\rm DM}^2 \left( s, \Omega \right) \dd s,
\end{equation}
where $s$ is a radial coordinate, and the $D$ factor
\begin{equation}
    \label{eq:D def}
    \frac{\dd D}{\dd \Omega} = \int  \rho_{\rm DM} \left( s, \Omega \right) \dd s,
\end{equation}
we arrive at the total photon fluxes per unit solid angle
\begin{align}
     \frac{\dd^2\Phi^{\rm ann}}{\dd E_\gamma \dd \Omega} &= \frac{\dd \Phi_{\rm pp}^{\rm ann}}{\dd E_\gamma} \frac{\dd J}{\dd \Omega}, \\
     \frac{\dd^2\Phi^{\rm dec}}{\dd E_\gamma \dd \Omega} &= \frac{\dd \Phi_{\rm pp}^{\rm dec}}{\dd E_\gamma} \frac{\dd D}{\dd \Omega},
\end{align}
where we note that we have assumed that the cosmological redshift variation across the source is negligible, so we can factor out \cref{eq:Phi_pp ann,eq:Phi_pp dec} from the line of sight integral.

\section{Gamma-ray data}
\label{sec:Observational data}

In this work we use gamma-ray observations from \textit{Fermi}-LAT and analyse these using the Fermi Tools\footnote{\url{https://fermi.gsfc.nasa.gov/ssc/data/analysis/software/}} and \fermipy \citep{FermiPy}. To mitigate the effect of contamination from cosmic rays we consider photons within the event class SOURCEVETO. We select all photons in this event class of energy $500{\rm \, MeV} - 50 {\rm \, GeV}$ between mission weeks 9 and 634 which are flagged as belonging to the upper quartile of angular resolution (PSF3) and set the maximum zenith angle to be $90^\circ$. We subdivide these data into nine logarithmically space energy bins, and then bin spatially onto \healpix\footnote{\url{http://healpix.sf.net}} \citep{Zonca_2019,Gorski_2005} maps. Although the angular resolution of the data corresponds to $\nside \approx 1024$, we compare it to the theoretical maps at $\nside = 256$ for computational efficiency. In \cref{sec:Sytematics JD maps} we find that out results are not significantly affected by this choice.

Because of the high density and close proximity of the centre of our own galaxy, one would expect that a \ac{DM} annihilation or decay signal would be dominated by this region. However, the constraint or detection one would obtain from studying this region would be sensitive to the modelling of the Milky Way density profile, and one would have to ensure that such a signal could not arise due to potentially incorrect modelling of the galactic diffuse or isotropic components, or through processes not captured by these models, such as an unresolved population of millisecond pulsars. This is a complicated yet feasible task \cite[see e.g.][]{Cholis_2021}, but is beyond the scope of this work; here we wish to produce constraints on \ac{DM} annihilation and decay that are independent of the GCE so we simply mask the galactic plane, with the aim that any constraint or detection is driven by the density fields inferred in \cref{sec:density_inference}. We therefore mask the region with galactic latitude $|\lambda|<30^\circ$.

\section{Methods}
\label{sec:Methods}

In this section we detail how we construct the full-sky templates for \ac{DM} annihilation and decay and how these are compared to the gamma-ray data to constrain the annihilation cross section and decay rate. In \cref{sec:density_inference} we describe the constrained simulations used to generate these templates and in \cref{sec:DM maps} we describe how the $J$ and $D$ factors are computed from the \ac{DM} particles in these simulations. These templates are combined with those from \cref{sec:Non-DM Templates} that describe non-\ac{DM} contributions to the gamma-ray sky, and we compare these to the data using the likelihood model in \cref{sec:Likelihood}.

\subsection{Bayesian large-scale structure inference}
\label{sec:density_inference}

To compute the $J$ and $D$ factors, we use the results of applying the \borg algorithm \citep[see e.g.][]{BORG_1,BORG_2,BORG_3,BORG_4,Lavaux_2016} to the 2M++ galaxy compilation \citep{Lavaux_2016, Jasche_Lavaux}. The \borg algorithm applies a Bayesian forward model for the observed number densities of galaxies to infer both the present-day three-dimensional density field and the corresponding \ac{ICs}. The algorithm incorporates a physical model for gravitational structure formation and marginalises over galaxy bias parameters.

Specifically, we use a set of \ac{DM}-only constrained simulations, dubbed \csiborg \citep{Bartlett_2021_VS,antihalos,Hutt_2022}. These start by taking 101 sets of $z=69$ \ac{ICs} from across the \borg chain, covering a box length of $677.7\Mpch$ with $256^3$ voxels. Inside a sphere of radius $155\Mpch$ centred on the Milky Way, these \ac{ICs} are augmented with white noise to a resolution of $2048^3$, resulting in a particle mass of $4.38 \times 10^9\;\Msun$. These simulations are then run to $z=0$ using \texttt{RAMSES} \citep{ramses}, giving 101 $N$-body realisations of the local \ac{DM} distribution. Only the central, high-resolution region is refined; the rest of the \borg box is retained only to include the effect of longer-wavelength modes. Both \borg and \csiborg adopt the cosmology $T_\text{CMB}=2.728$ K, $\Omega_\text{m} = 0.307$, $\Omega_\Lambda = 0.693$, $\Omega_\text{b} = 0.04825$, $H_0 = 70.5$ km s$^{-1}$ Mpc$^{-1}$, $\sigma_8 = 0.8288$ and $n_{\rm s}=0.9611$. Our results are therefore conditioned on this cosmology; a study of the cosmology dependence of our results is beyond the scope of this work.

To improve the effective resolution of our calculation in high-density regions we run a halo finder on the \csiborg particles and use analytic formulae for a specified density profile. Specifically, we use the watershed halo finder \texttt{PHEW}  \citep{PHEW} which runs on the fly as a patch to \texttt{RAMSES}. This splits the negative density field into basins that share a common local minimum via steepest descent. These basins are then combined according to user-defined thresholds on the density of saddle points between basins to merge low-mass subhalos into their parents; we use the standard threshold value from \citet{PHEW} of $200\rho_c$. The halo catalogues are publicly available for the full \csiborg suite \citep{max_zenodo}.

\subsection{Computing the \texorpdfstring{$J$}{J} and \texorpdfstring{$D$}{D} factors}
\label{sec:DM maps}

\begin{figure*}
\centering
\begin{minipage}{.45\textwidth}
  \centering
  \includegraphics[width=\linewidth]{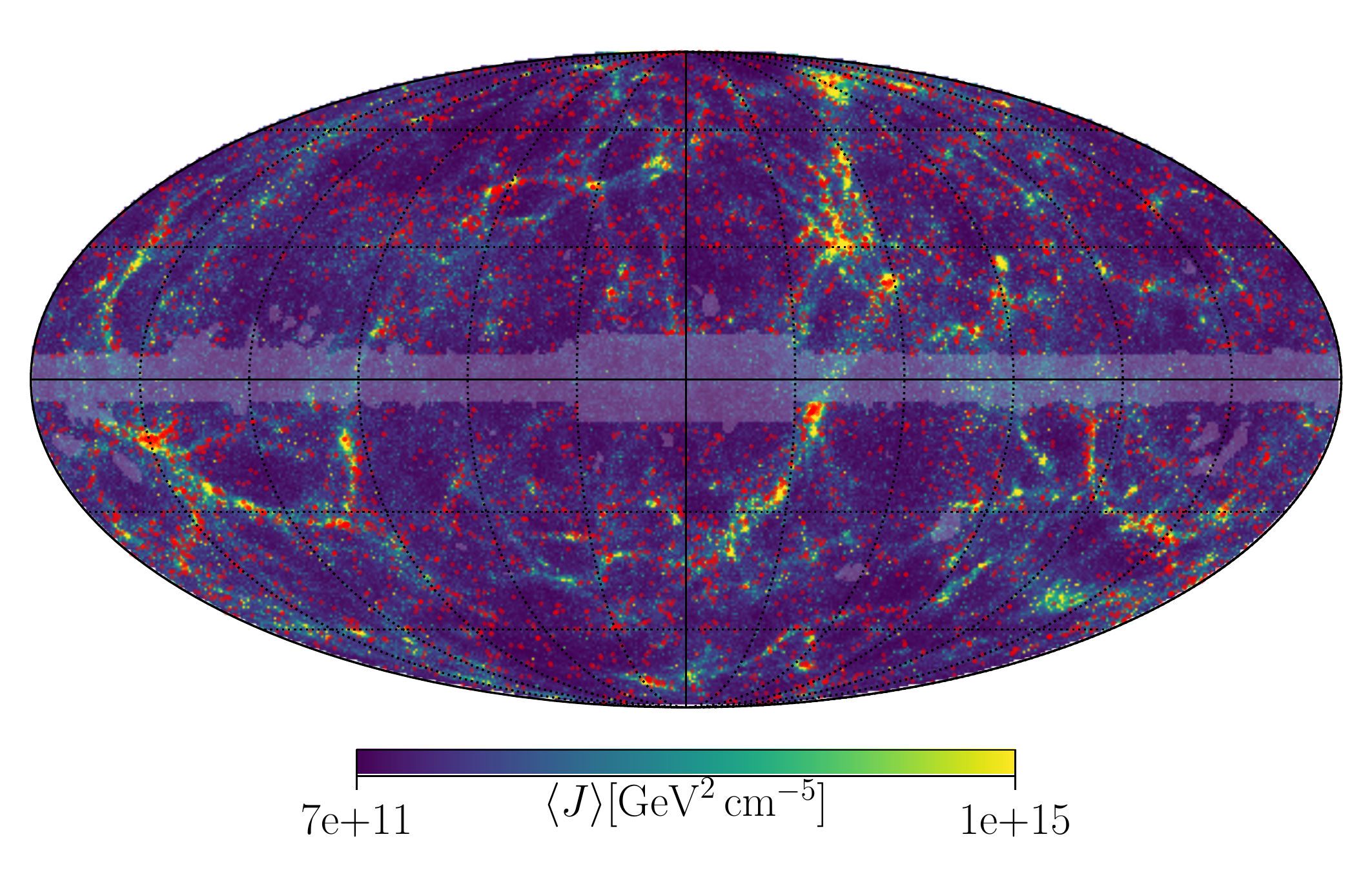}
\end{minipage}
\begin{minipage}{.45\textwidth}
  \centering
  \includegraphics[width=\linewidth]{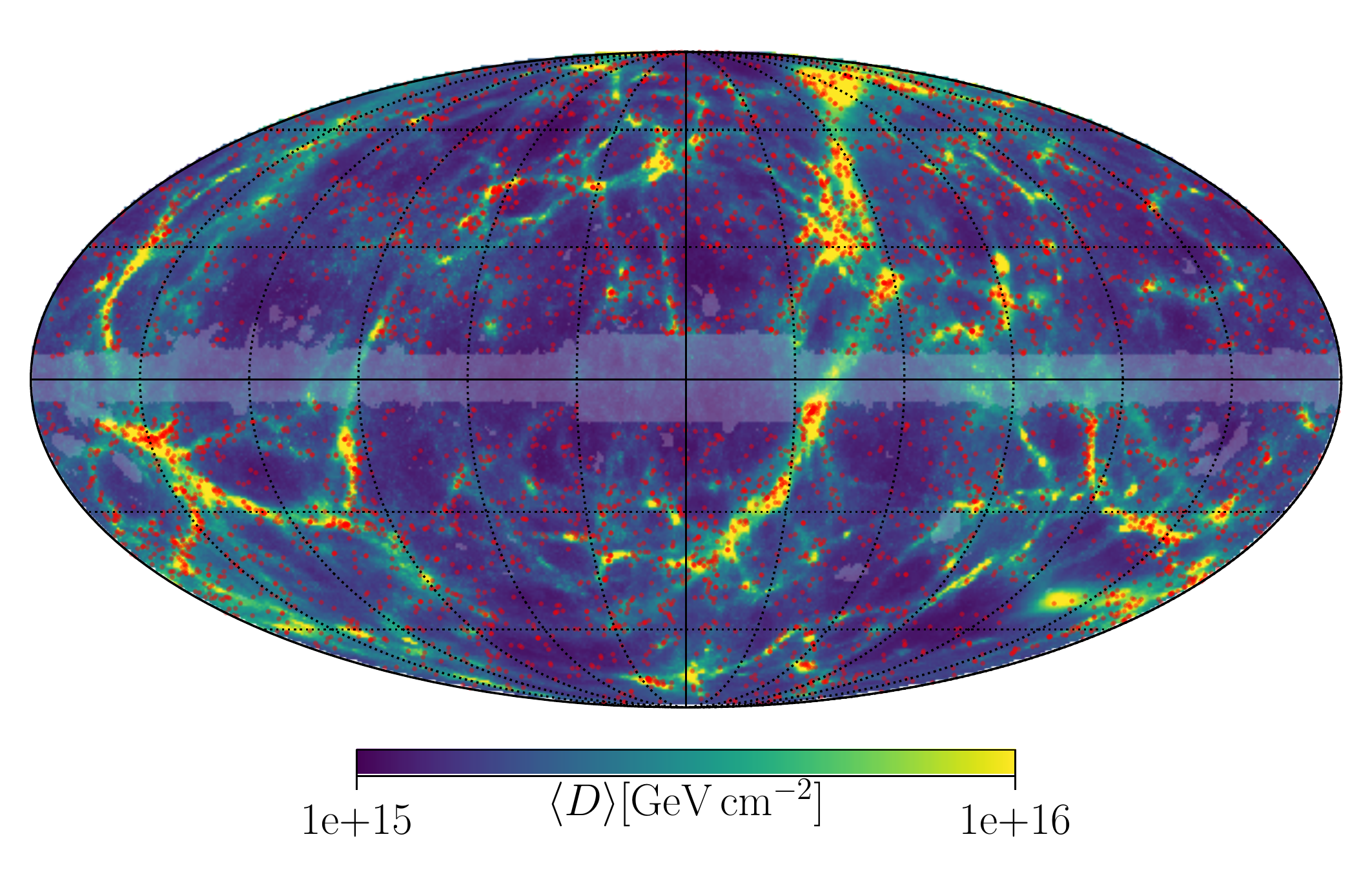}
\end{minipage}
\caption{
Mollweide projection in galactic coordinates of the ensemble mean $J$ and $D$ factors over the \csiborg realisations, alongside the brightest $\sim$ 6000 galaxies from the 2M++ dataset (red points). One can see that the galaxy number density is higher in the regions of large $J$ and $D$ factor, i.e. at the peaks of the underlying \ac{DM} distribution. We overplot the mask on completeness used in the \borg inference of the initial conditions \citep{Lavaux_2016, Jasche_Lavaux} (faded region near the galactic plane masked out).}
\label{fig:jd_ensemble}
\end{figure*}

The halo finding allows us to split \csiborg particles into two types: those that belong to halos and those that do not. Since the $J$ factor depends on the square of the density, it is more sensitive to the small-scale matter distribution, and thus we must treat halos separately from the background density field in order to account for structures below the resolution of the \csiborg simulations. The $D$ factor is less sensitive to these small scales, and thus we treat all particles equally in this case. $D$ is computed using the procedure outlined in \cref{sec:Smoothed density field}, where we use all particles. We compute $J$ as in \cref{sec:Smoothed density field}, but only considering nonhalo particles, and add this to the contribution from halos, which are treated as in \cref{sec:Halos}. We plot the resulting ensemble mean $J$ and $D$ factor maps in \cref{fig:jd_ensemble}.

For the $J$ factor maps, we generate the templates on a higher resolution \healpix grid than that on which we perform the inference ($\nside=2048$ instead of $\nside=256$) and subsequently degrade them. Because of the nonlinear dependence of $J$ on $\rho$, this allows for a more faithful representation of the density field than if $J$ was initially calculated at $\nside=256$, which is especially important for the regions of the sky corresponding to halos produced in \csiborg . We concluded the \healpix resolution of $\nside=2048$ was sufficient by comparing the total $J$ factor to those with increasing \healpix resolution ($\nside=4096, 8192$) since the change in total $J$ factor was at the subpercent level with increasing \nside.

Regarding the $D$ factor calculation, we directly calculate the line-of-sight integral of the density field within a given \healpix pixel at the selected resolution. The convergence of the total all-sky $D$ factor with this procedure is, of course, present for all considered resolutions, since it should be simply proportional to the total mass within the \csiborg volume.

\subsubsection{Smoothed density field}
\label{sec:Smoothed density field}

We wish to determine the density of \ac{DM} particles that do not belong to a halo on a regular Cartesian grid with $N_{\rm grid}=1024$ grid points per side. To do this, we adopt a procedure based on \ac{SPH} \citep{Monaghan_1992} as described in \citep{Colombi_2007} and outlined below. Using the \ac{SPH} algorithm over, e.g., a \ac{CIC} approach allows us to better capture the peaks of the matter density field, since the \ac{SPH} kernel will adapt to the local density of matter, in contrast to the \ac{CIC} approach which has a fixed kernel corresponding to a trilinear interpolation scheme.
We compare the results of using a \ac{SPH} kernel to a \ac{CIC} algorithm in \cref{sec:Sytematics JD maps}.

First, we determine the number of particles, $N_{\rm p}$, within the cell corresponding to each grid point $(i,j,k)$. We then define
\begin{equation}
    N_{\rm X} = \max \left( N_{\rm p}, N_{\rm SPH} \right),
\end{equation}
where $N_{\rm SPH}=32$. The choice for this number of neighbours is partly motivated by the typical number of edges linking a node to its neighbours in a Delaunay tesselation. That number is approximately 16 for a Euclidean three-dimensional vector space \citep{Okabe2000,Neyrinck2008}. We pick a value twice as big as we intend the filter to have a larger reach than the first layer of neighbours. We then find the mass associated with this grid point by considering the nearest $N_{\rm X}$ particles to be
\begin{equation}
    \tilde{m}_{ijk} =
    \frac{1}{R_{ijk}^3}
    \sum_{l=1}^{N_{\rm X} -1} m_l W_l \mathcal{S} \left(\frac{d_l}{R_{ijk}} \right),
\end{equation}
where $R_{ijk}$ is half the distance to the farthest of the $N_{\rm X}$ particles from the grid point, $m_l$ is the mass of particle $l$, which is at a distance $d_l$ from the grid point, $W_l$ is the weight for particle $l$,
\begin{equation}
    W_l = \left( \sum_{ijk} \frac{1}{R_{ijk}^3} \mathcal{S} \left(\frac{d_l}{R_{ijk}} \right) \right)^{-1},
\end{equation}
and the interpolating function, $\mathcal{S}$, is chosen to be
\begin{equation}
	 \mathcal{S} \left( x \right) = 
		\begin{cases}
			\mbox{$1 - \frac{3}{2} x + \frac{3}{4}x^3$}, & 0 \leq x < 1 \\
			\mbox{$\frac{1}{4} \left( 2 - x \right)^3$}, & 1 \leq x < 2 \\
			\mbox{$0$}, & \text{otherwise}.
		\end{cases}
\end{equation}
If the spacing between grid points is $\Delta r$, then the density assigned to each site is
\begin{equation}
    \tilde{\rho}_{ijk} = \frac{\tilde{m}_{ijk}}{\Delta r^3}.
\end{equation}
To compute the $J$ and $D$ factors, we compute \cref{eq:J def,eq:D def}, respectively, along the line of sight corresponding to each \healpix pixel at the chosen resolution. We integrate up to the edge of the simulated volume and perform trilinear interpolation of the density field onto the line of sight. The convergence of this approach was checked by increasing the resolution of the \ac{SPH} kernel. The total assigned mass to the grid was consistent among all resolutions we tried ($N_{\rm grid} = 256$, $512$, $1024$); therefore we opted for $N_{\rm grid} = 1024$ for our final $J$ and $D$ factor calculations. 

\subsubsection{\texorpdfstring{$J$}{J} factor from halos}
\label{sec:Halos}

To include the contribution from particles inside halos, we use a custom extension of the \clumpy package\footnote{\url{https://clumpy.gitlab.io/CLUMPY}} \citep{Charbonnier_2012,Bonnivard_2016,Hutten_2019}. In this section we review how the $J$ factor is calculated in this package, as well as our assumptions for the halo density profiles. \par
We assume that all halos are spherically symmetric and that the total density profile, $\rho_{\rm tot}$, can be described by a simple analytic form. For our fiducial case, we consider the three-parameter family of profiles \citep{Hernquist_1990,Zhao_1996}
\begin{equation}
    \rho_{\alpha\beta\gamma} \left( r \right) = \frac{2^{\frac{\beta - \gamma}{\alpha}}\rho_{\rm s}}{\left(r/r_{\rm s} \right)^\gamma \left( 1 + \left( r / r_{\rm s} \right)^\alpha \right)^{\frac{\beta - \gamma}{\alpha}}},
    \label{eq: ZHAOprofile}
\end{equation}
where $r_{\rm s}$ is the scale radius, $\rho_{\rm s}$ is the density at $r_{\rm s}$, and $\alpha$ describes the sharpness of the transition between the inner ($\gamma$) and outer ($\beta$) logarithmic slopes. For a NFW profile, $\alpha=1$, $\beta=3$, $\gamma=1$. In \cref{sec:Sytematics JD maps}, we also consider the Einasto profile \citep{Navarro_2004,Springel_2008}
\begin{equation}
    \rho_{\rm EINASTO} \left( r \right) = \rho_{-2} \exp \left( - \frac{2}{\alpha} \left( \left( \frac{r}{r_{-2}} \right)^\alpha -1 \right) \right).
    \label{eq: EINASTOprofile}
\end{equation}
In the above, $r_{-2}$ represents the radius at which the logarithmic slope of the profile equals $-2$, while $\rho_{-2} \equiv \rho(r_{-2})$. For both profiles we calculate the parameters defining the profile from the total halo mass and corresponding concentration. The mass-concentration relation we use is shown in \cref{eq: mass_concentration_relation}. We also note that the parameter $\alpha$ of the Einasto profile is determined as a function of virial peak height \citep{2008MNRAS.387..536G}. Given that all of our halos are at very low redshift ($z \lesssim 0.05$) and almost all have masses within $M \in [10^{12}, 10^{15}]\, \Msun$, the parametric relation is quite accurate. We use the \textsc{colossus} package throughout \citep{diemer2018colossus}. We note that, since the Einasto parameters are fitted to halos produced in \ac{DM}-only simulations, this captures only the uncertainty in $N$-body modelling. Baryons induce a potentially larger effect, which is, however, harder to model reliably. We discuss this further in \cref{sec:Systematics Halo Profile}.

One would also expect that a halo contains a large number of ``clumps'' or subhalos, such that the true smooth component of the density profile is \citep{Pieri_2011}
\begin{equation}
    \rho_{\rm sm} \left( r \right) = \rho_{\rm tot} \left( r \right) - \langle \rho_{\rm subs} \left( r \right) \rangle,
\end{equation}
where $\langle \rho_{\rm subs} (r) \rangle$ gives the average contribution from the substructure. If clump $i$ has density profile $\rho_{\rm cl}^i$, then it contributes to the total $J$ factor value within a \healpix pixel, $p$, as
\begin{align}
        J_p &= \int_{p, \Delta_{\rm{halo}}} \left( \rho_{\rm sm}(s,\Omega) + \sum_i \rho_{\rm cl}^{i}(s,\Omega)\right)^2 \dd s \dd\Omega \nonumber\\
        & = J_{{\rm sm}, p} + J_{{\rm subs}, p} + J_{{\rm cross}, p},
\end{align}
where
\begin{align}
    J_{{\rm sm}, p} &= \int_{p, \Delta_{\rm{halo}}} \rho_{\rm sm}^2(s,\Omega)  \dd s \dd \Omega,\\
    J_{{\rm subs}, p} &= \int_{p, \Delta_{\rm{halo}}} \left(  \sum_i \rho_{\rm cl}^{i}(s,\Omega) \right)^2 \dd s \dd \Omega, \label{eq:Jsubs definition} \\
    J_{{\rm cross}, p} &= 2 \int_{p, \Delta_{\rm{halo}}} \rho_{\rm sm}(s,\Omega) \left( \sum_i \rho_{\rm cl}^{i}(s,\Omega) \right) \dd s \dd \Omega,
\end{align}
and $\Delta_{\rm{halo}}$ represents the intersection of the halo volume with the cone spanned by the pixel $p$. Our task is therefore to determine the distribution of clumps for a given halo and to calculate these integrals. Here, we readily use the solution provided by the \clumpy package and describe it briefly below.

Assuming that a given halo has $N_{\rm tot}$ independent clumps, we factorise the distribution for the number of clumps with some mass $M$, concentration $c$, in some region $\dd V = \dd^3 r$ as \citep{Lavalle_2008,Bonnivard_2016}
\begin{equation}
    \frac{\dd N}{\dd V \dd M \dd c} =
    N_{\rm tot} \frac{\dd \mathcal{P}_V \left( r \right)}{\dd V}
    \frac{\dd \mathcal{P}_M \left( M \right)}{\dd M}
    \frac{\dd \mathcal{P}_c \left( M, c \right)}{\dd c}.
    \label{eqn: clumps_pdf}
\end{equation}
Since the clumps form before the host halos within $\Lambda$CDM, their spatial distribution will follow the host \ac{DM} density profile. This has been shown to be a good assumption in simulations of galaxy-sized halos \citep{Springel_2008, ludlow2009unorthodox}. Given the self-similar nature of collapse of $\Lambda$CDM halos, we extrapolate this conclusion to halos from our \csiborg ensemble. We assume that the distribution of masses is a power law
\begin{equation}
    \frac{\dd \mathcal{P}_M }{\dd M} \propto M^{- \alpha_{M}},
\end{equation}
in the range $M \in [10^{-6}M_{\sun}, 10^{-2} M_{\rm h}]$ for a halo of mass $M_{\rm h}$, where $\alpha_{M} = 1.9$ (see Sec. 2.3 of \citep{hutten2016dark} and references therein). 
Again, the values are motivated by numerical simulations of Milky Way sized halos, which we extrapolate to bigger halos present in our forward model.

Besides modelling the uncertainty due to the spatial and mass distribution of substructure, the \clumpy package also allows us to include the uncertainty in the mass-concentration relation. For the substructure component, we consider two cases for the concentration distribution. In the first case, we assume that the concentration of all substructure halos is a deterministic function of the mass
\begin{equation}
    \frac{\dd \mathcal{P}_c }{\dd c} = \delta \left( c - \bar{c} \left( M \right) \right),
\end{equation}
where we define $\delta$ as the Dirac-delta distribution. The second possibility that we follow is that the concentration is log normally distributed about this mean
\begin{equation}
    \frac{\dd \mathcal{P}_c }{\dd c} =
    \frac{1}{\sqrt{2 \pi} c \sigma_c \left( M \right)}
    \exp \left( - \frac{\left( \log c - \log \left( \bar{c}\left(M \right) \right) \right)^2}{2 \sigma_c^2} \right).
    \label{eq: mass_concentration_dpcdc}
\end{equation}
Motivated by \citep{Fornasa:2016ohl}, the substructure halos are assumed to have the following mass-concentration relation \citep{Sanchez_Conde_2014}
\begin{equation}
    \bar{c} \left( M \right) = \sum_{j=0}^5 C_j \left[ \ln \left( \frac{M}{h^{-1} M_{\sun}}\right) \right]^j,
    \label{eq: mass_concentration_relation}
\end{equation}
where $C_j = (37.5153, -1.5093, 1.636 \times 10^{-2}, 3.66 \times 10^{-4}, -2.8927 \times 10^{-5}, 5.32 \times 10^{-7})$, with $\sigma_{\rm c}=0.0$; i.e., we assume all substructure halos of the same mass have the same concentration. In \cref{sec:Systematics substructure uncertainties} we consider a nonzero value $\sigma_{\rm c} = 0.2$ in \cref{eq: mass_concentration_dpcdc}, as motivated by \citep{bullock2001profiles, sanchez2014flattening, wechsler2002concentrations} as a comparison.

\begin{figure}
\centering
    \subfloat[]{
         \centering
         \includegraphics[width=\columnwidth]{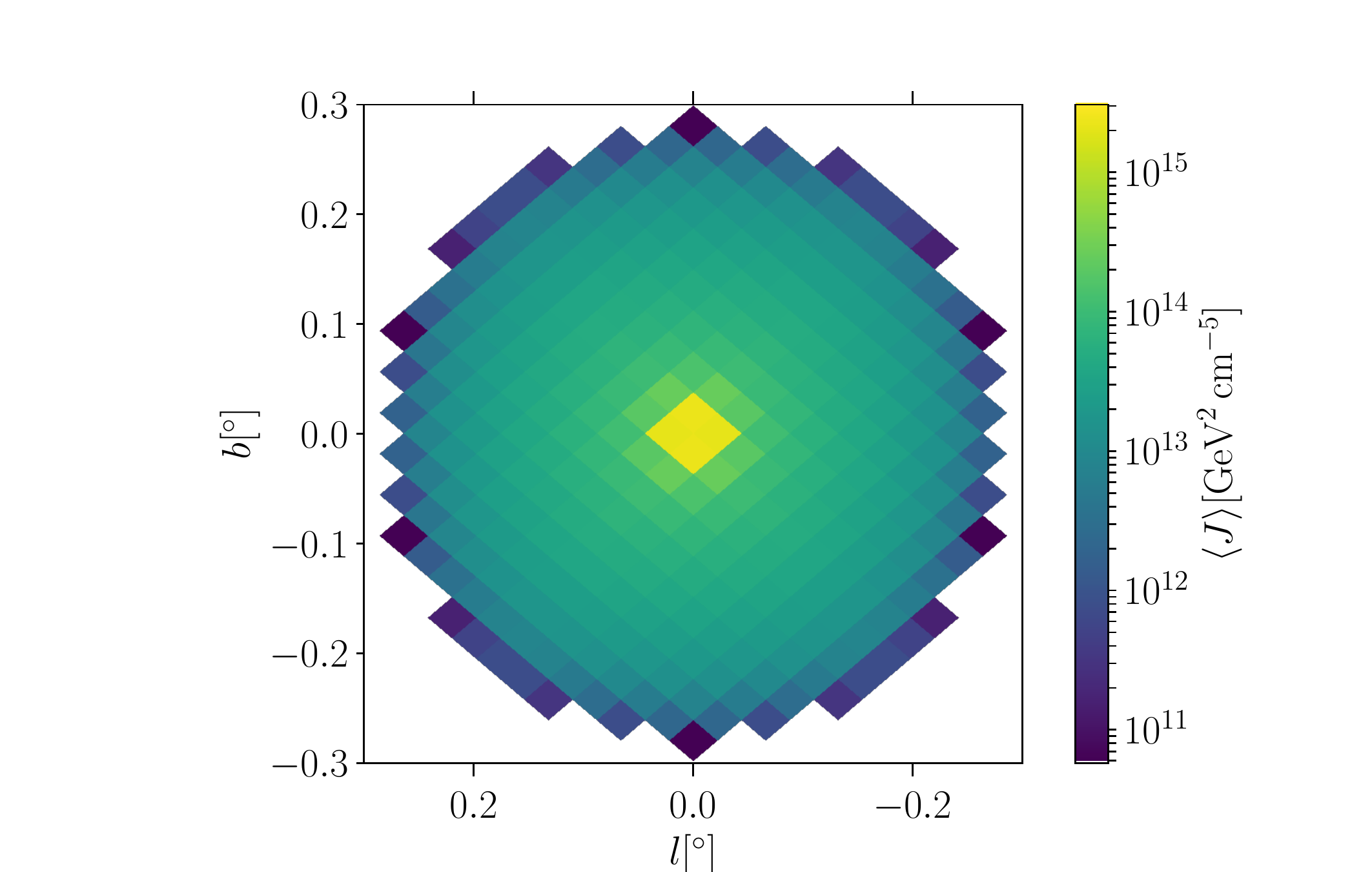}
         \label{subfig:NFW_J}
     }\\ \vspace*{-\baselineskip}
     \subfloat[]{
         \centering
         \includegraphics[width=\columnwidth]{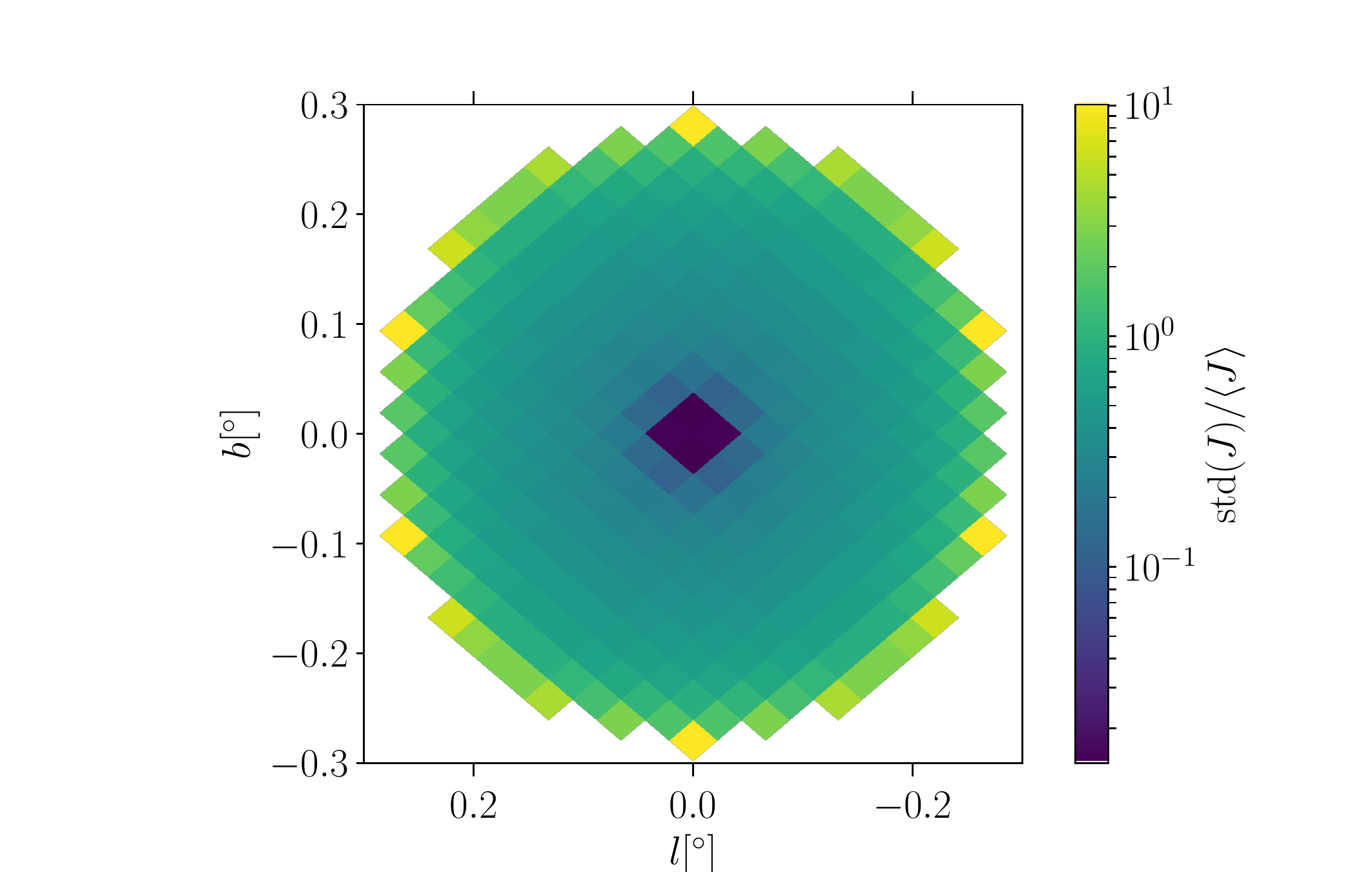}
         \label{subfig:NFW_J_std}
     }
    \caption{\protect\subref{subfig:NFW_J} $J$ factor of a typical NFW halo within \csiborg ($M_{\rm h} \approx 5\times 10^{13} \, M_{\odot}$) and \protect\subref{subfig:NFW_J_std} the corresponding relative fluctuations in the $J$ factor due to the substructure contribution. The quantities $\langle J \rangle$ and $\mathrm{std}(J)$ are calculated according to \cref{eqn: mean_J_p,eqn: std_J_p} respectively. As can be seen, the relative size of fluctuations in the $J$ factor grows toward the outskirts. Qualitatively similar features are observed if we assume an Einasto profile. Note that here we placed the halo at the centre of the \healpix grid for numerical convenience.
    }
    \label{fig: J_factor_typical_halo_and_stdJ}
\end{figure}

Given that we do not resolve substructures of all the halos present with our simulation, we assume the resulting distribution for $J_{\rm sm}$ and $J_{\rm subs}$ to be a Gaussian. Therefore, we only need to find the mean and variance of these contributions in each \healpix pixel. We define the 1-clump luminosity to be
\begin{equation}
    L \left(M, c \right) \equiv \int_{V_{\rm{subhalo}}} \rho_{\rm cl}^2 \left( r ; M,c \right) \dd^3 r,
\end{equation}
and its moments as
\begin{equation}
    \left< L^n \right> \equiv
    \int_{M_1}^{M_2} \frac{\dd \mathcal{P}_M }{\dd M}
    \int \frac{\dd \mathcal{P}_c }{\dd c}
    L^n \dd c \, \dd M,
    \label{eqn: luminosity_moments}
\end{equation}
for a given mass range of clumps $[M_1, M_2]$, while the mean contribution of $J_{{\rm subs},p}$ from this volume is
\begin{equation}
    \left< J_{{\rm subs},p} \right>
    = N_{\rm tot} \int_{p, \mathcal{V}_{\ell}} \frac{\dd \mathcal{P}_V}{\dd V} \ell^{-2} \dd V\left< L \right>,
\end{equation}
with $N_{\rm{tot}}$ representing the total number of clumps within the selected mass range $[M_1, M_2]$. Note that we assume that the clumps are nonoverlapping, such that the cross terms in \cref{eq:Jsubs definition} can be neglected. For more details on how these quantities are defined we refer the reader to the \clumpy related publications \citep{Charbonnier_2012, Bonnivard_2016, Hutten_2019}.

Note that since we are assuming unresolved substructures for our \csiborg extragalactic halos, we are integrating over the total subhalo volume $V_{\rm{subhalo}}$ for the subhalo luminosity. Furthermore, since there can be many subhalos present within the line of sight determined by the given \healpix pixel we are also accounting for the span of the host halo along this line of sight through $\mathcal{V}_{\ell} \equiv [\ell_{\rm{min}}, \ell_{\rm{max}}]$, with $\ell_{\rm{min}}$ and $\ell_{\rm{max}}$ being the closest and farthest points of the host halo along this line of sight. Since these integrals do not have a closed form for general \ac{DM} profiles, we evaluate all numerically.

Given that the mean of some power of the distance from the observer, $\ell$, to a clump which falls inside the \healpix pixel $p$ is
\begin{equation}
    \left< \ell^n_p \right> =
    \int_{p, \mathcal{V}_{\ell}} \ell^{n+2} \frac{\dd \mathcal{P}_V }{\dd V} \dd \ell \, \dd \Omega,
\end{equation}
we can write the variance on $J_{\rm{subs}}$ as
\begin{equation}
    \sigma^2_{{J_{{\rm subs}, p}}} = 
    \left< L^2 \right>  \left< \ell^{-4}_p \right>  - \left< L \right>^2  \left< \ell^{-2}_p \right>^2,
\end{equation}
since $L$ and $\ell$ are independent.
For the cross term, $J_{{\rm cross}, p}$, we use that its mean is
\begin{equation}
    \left< J_{{\rm cross}, p} \right>
    = 2 \int_{p, \mathcal{V}_{\ell}} \rho_{\rm sm} \left< \rho_{\rm subs} \right> \dd \ell \, \dd \Omega,
\end{equation}
while its variance can be computed as
\begin{equation}
    \sigma_{J_{{\rm cross}, p}}^2 = 4\int_{p, \mathcal{V}_{\ell}} \rho_{\rm{sm}}^2(l,\Omega) 
    \sigma_{\rm{subs}}^2(l,\Omega) \dd \ell \, \dd \Omega,
    \label{eq: sigma_J_cross}
\end{equation}
with
\begin{equation}
    \sigma_{\rm{subs},p}^2 \equiv \sigma_{\rm{subs}}^2(l,\Omega) = 
    \langle \rho_{\rm{subs}}^2(l,\Omega) \rangle
    -
    \langle \rho_{\rm{subs}}(l,\Omega) \rangle^2,
\end{equation}
\begin{equation}
    \begin{split}
        \hspace{-5mm}        
        \langle \rho_{\rm{subs}}(l,\Delta \Omega) \rangle =  &\int_{\Delta \Omega, \mathcal{V}_{\ell}} \frac{\dd \mathcal{P}_V }{\dd V}
        \int_{\mathcal{V}_{M}} \frac{\dd \mathcal{P}_M }{\dd M}
        \int_{\mathcal{V}_{c(M)}} \frac{\dd \mathcal{P}_c }{\dd c} \\
        &\times \rho_{\rm{subs}}(\ell, \Omega ; M, c(M)) \, \dd \ell \dd \Omega \dd M \dd c,
    \end{split}
\end{equation}
with $\mathcal{V}_{M}$ and $\mathcal{V}_{c(M)}$ representing the mass and corresponding concentration ranges for the subhalo distribution respectively.

We decide to include only $\sigma_{J_{{\rm subs},p}}$ as it is the dominant source of uncertainty. This can intuitively be understood from the \cref{eq: sigma_J_cross}. We note that the integrand is negligible both in the outskirts of the host halo, since $\rho_{\rm sm} \sim 0$ and $\sigma_{\rm subs}^2$ remains finite, and in the very centre of the host halo, since $\sigma_{{\rm subs}, p} \sim 0$. Therefore, $J_{{\rm cross}, p}$ contributes only at a very limited range of scales. Furthermore, for our halos, there is a clear hierarchy between the cross and subs term $J_{{\rm cross}, p} \lesssim 0.01-0.1 J_{{\rm subs}, p}$; therefore, we focus only on the $\sigma_{J_{{\rm subs}, p}}$ as the dominant source of uncertainty of the $J$ factor due to substructure. 

We hence write that the distribution followed by the $J$ factor for a given halo in a given pixel is given by a Gaussian with mean
\begin{equation}
    \left< J_p \right> = J_{{\rm sm}, p} + \left< J_{{\rm subs}, p} \right> + \left< J_{{\rm cross}, p} \right>,
    \label{eqn: mean_J_p}
\end{equation}
and variance
\begin{equation}
    \sigma^2_p = \sigma_{J_{{\rm subs}, p}}^2 + \sigma_{J_{{\rm cross}, p}}^2 \approx \sigma_{J_{{\rm subs}, p}}^2.
    \label{eqn: std_J_p}
\end{equation}
We calculate these quantities for all \csiborg realisations. To distinguish between these, we introduce a second index, $j$, to label the simulation, i.e. $\left< J_{p j} \right> $ is $\left< J_p \right> $ for \csiborg simulation $j$, and likewise $\sigma_{p j}$ is $\sigma_p$ for the same simulation.

In \cref{fig: J_factor_typical_halo_and_stdJ}, we show the result of the model for a typical halo within \csiborg with a mass of $M_h \approx 5 \times 10^{13}\,\Msun$. We see that the effects from the term in $\sigma^2_{J_{\rm{subs}}}$ cannot be neglected, especially in the outskirts of the halo. In the very centre, where the structure of the halo is dominated by the smooth component, the fluctuations in the $J$ factor due to the substructure are negligible, amounting to only few percent, while in the outer parts these fluctuations become more important. This is an expected result given that the boost in $J$ factor due to substructure becomes more important in the outer edges, where the smooth component, $J_{\mathrm{sm},p}$, is subdominant with respect to the substructure $J$ factor, $J_{\mathrm{subs},p}$. This behaviour is identical for an Einasto profile.

Besides this, we also include the contributions of sub-subclumps to the $J$ factor of halos, using one additional level of substructure, which is the default setting of the \clumpy code. Because of the increased computational cost, we considered a two-level substructure contribution for our halos for only one \csiborg realisation. Including additional substructure levels will result in an overall change in the $J$ factor of less than $\sim 5\%$ \citep{Bonnivard_2016}, and ignoring such levels will make our constraints on $\sigv$ conservative since this will systematically underestimate the $J$ factor.

In conclusion, to obtain the total all-sky $J$ factor, we combine the line-of-sight calculation for the density field obtained from particles outside of halos detected within \csiborg realisations with the component coming from the halo particles of the \csiborg by treating the halos as presented in this section, utilising the \clumpy code. This final template is used in the inference pipeline. We discuss the numerical convergence of these calculations in \cref{subsec: systematic_uncertainties}.

\subsection{Non-\ac{DM} templates}
\label{sec:Non-DM Templates}

\begin{figure}
     \centering
     \includegraphics[width=\columnwidth]{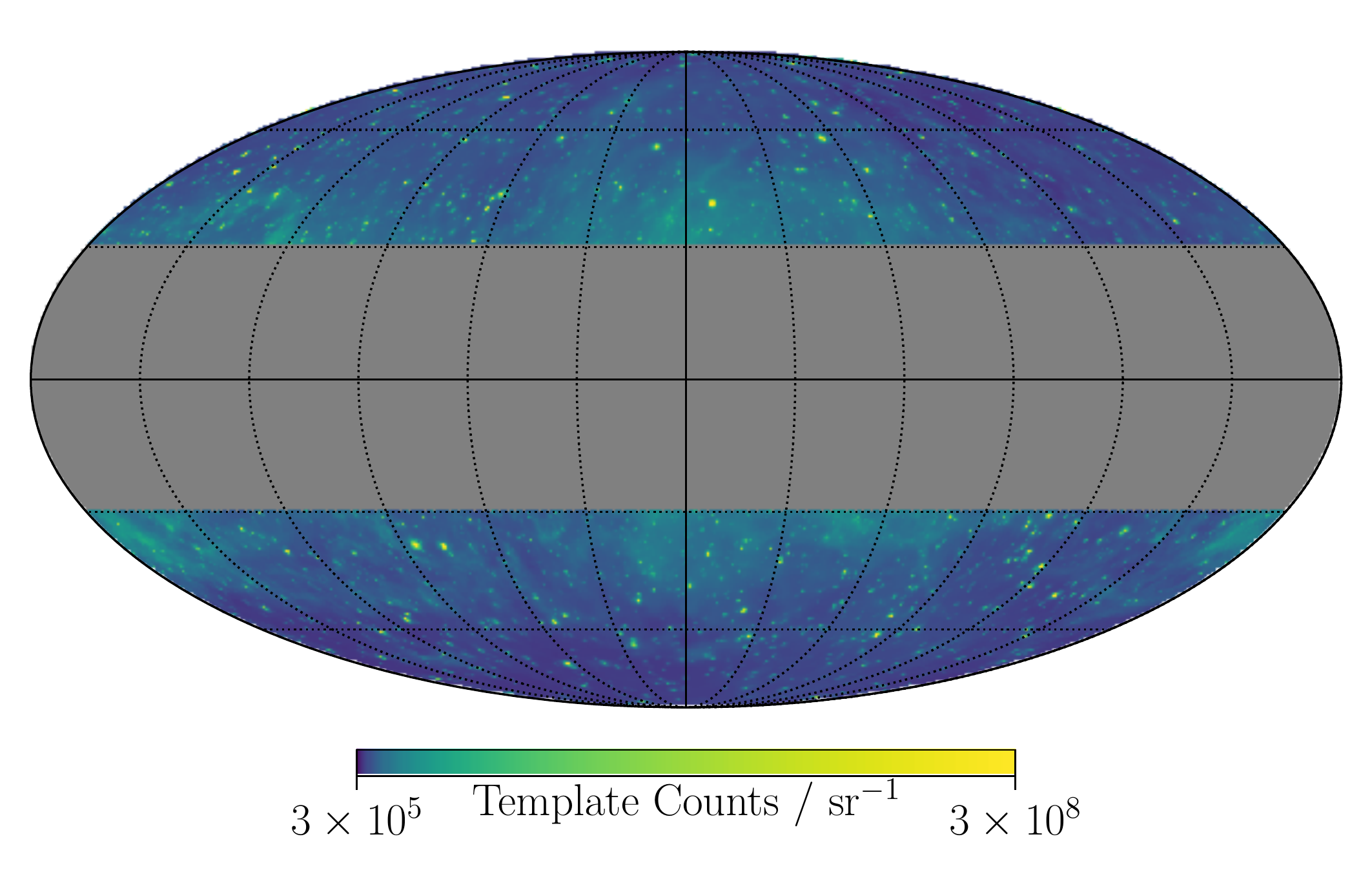}
    \caption{Mollweide projection in galactic coordinates of the template predictions across the full energy range considered in this work. For visualisation, we sum the isotropic, galactic diffuse, and point source templates assuming each has a unit amplitude. In our inference we simultaneously infer the normalisation of each of the three components, with a different amplitude for each energy bin, and the contribution proportional to the $J$ or $D$ factor.
    }
	\label{fig:Components_used_for_inference}
\end{figure}

To constrain the parameters describing \ac{DM} annihilation or decay, one also needs to take into account other sources of gamma rays. We consider a model with three contributions: our own galaxy (gal), an isotropic background (iso) and point sources (psc). We produce separate templates, $\{T_{i}^t \left(\hat{r}\right)\}, \, t \in \{ {\rm iso, \, gal, \, psc} \}$ for each energy bin, $i$, and assign each template a different normalisation, which we infer from the data. The sum of the three templates is plotted in \cref{fig:Components_used_for_inference}.

The isotropic component is designed to capture emission from unresolved extragalactic sources, residual cosmic rays and extragalactic diffuse sources. It consists of a spatially constant map, with a spectral shape given by the \textit{Fermi} Isotropic Spectral Template\footnote{\url{https://fermi.gsfc.nasa.gov/ssc/data/access/lat/BackgroundModels.html}}, but with an overall normalisation $A_i^{\rm iso}$, which we infer separately for each energy bin, $i$.

For the galactic component, we use the spatial models described in \citep{Acero_2016}, which are developed using spectral line surveys of HI and CO and infrared tracers of dust column density and a model of inverse Compton emission. These spatial templates describe the relative change in flux across different parts of the sky. We keep these fixed during our analysis and infer the normalisation, $A_i^{\rm gal}$, in each energy bin, which controls the total emission from our galaxy. We note that these templates are not designed to be used for analyses which aim to fit medium or large-scale diffuse structures, since the templates include a filtered residual map, which is smoothed to a few degrees. Any large-scale component not explicitly modelled when generating the templates will be absorbed by this residual and would be undetectable in our analysis. However, as can be seen in \cref{fig:jd_ensemble}, the $J$ and $D$ factor maps are dominated by small-scale features due to massive structures, and thus we are in the regime for which these templates are valid.

Finally, we produce a single template containing all point and extended sources from the models provided in the LAT 12-year Source Catalog (4FGL-DR3).\footnote{\url{https://heasarc.gsfc.nasa.gov/W3Browse/fermi/fermilpsc.html}} We introduce a free scaling parameter, $A_i^{\rm psc}$, for each energy bin which will be inferred. Since we find no cross-correlation between the point-source template and our own $J$ and $D$ factor templates, having a separate normalisation for the point-source template will not introduce new degeneracies (see \cref{sec:Systematics Non-DM}).

\subsection{Likelihood model}
\label{sec:Likelihood}

Instead of directly constraining the \ac{DM} annihilation or decay parameters, we split the inference into two parts. First, we assume that there is a contribution to the gamma-ray sky which is proportional to the $J$ or $D$ factor, i.e.\ for a given \csiborg simulation $j$, the flux in energy bin $i$ and pixel $p$, $\Phi_{i p j}$, has terms
\begin{equation}
    \Phi_{i p j} \supset \left( A_i^{\rm J} \frac{J_{p j}}{J_0} + A_i^{\rm D} \frac{D_{p j}}{D_0} \right) \mathcal{A}_p \Delta E_i,
\end{equation}
where $\Delta E_i$ is the width of the bin, $\mathcal{A}_p$ is the area of the pixel in steradians and $J_0$ and $D_0$ set the units. We choose $J_0 = 10^{13}  {\rm \, GeV^2\, cm^{-5}}$ and $D_0 = 10^{13} {\rm \, GeV\, cm^{-2}}$.
We fit for the total flux of such a contribution in each energy bin to obtain a spectrum. In the second half of the inference, we fit this spectrum to a series of models (including \ac{DM} annihilation and decay) in an attempt to determine the origin of such a signal.

This method has two main advantages. First, we can consider each energy bin and \csiborg simulation separately in the first half of the inference. Although this involves initially computing 909 \ac{MCMC} chains (one for each energy bin and for each \csiborg simulation), since the problem is embarrassingly parallelisable and because we only need to infer four or five parameters for each chain ($A_i^\text{J}$ and/or $A_i^\text{D}$, $A_i^\text{iso}$, $A_i^\text{gal}$, $A_i^\text{psc}$) compared to $\sim30$ if we combined the energy bins, we find that this approach is computationally more efficient. Second, by remaining agnostic to the origin of $A_i^{\rm J}$ or $A_i^{\rm D}$ until the second step, we are able to more easily determine which energy bins drive our constraints. Hence, it becomes simpler to compare different models since we do not need to rerun the map-level inference every time that we wish to change the \ac{DM} particle mass or decay channel (which is more computationally expensive).

\subsubsection{Inferring the spectrum}

We assume that photon counts in energy bin $i$ from the $J$ and $D$ factor contributions, as well as each of the contributions described in \cref{sec:Non-DM Templates} is Poisson distributed. The variation of the mean of the latter with sky position, $\hat{r}$, and energy is described by the known templates $\{T_{i}^t \left(\hat{r}\right)\}$, where $t$ labels the templates. For pixel $p$ and \csiborg realisation $j$, we then define
\begin{equation}
    J_{ipj} \equiv \frac{J_{p j}}{J_0} \Delta E_i, \quad 
    D_{ipj} \equiv \frac{D_{p j}}{D_0} \Delta E_i, \quad 
    T_{ip}^t \equiv \int_p T_{i}^t \left(\hat{r}\right) \dd \Omega,
\end{equation}
such that the mean number counts in pixel $p$ and energy bin $i$ is predicted to be
\begin{equation}
    \lambda_{ipj} = \mathcal{F}_{ip} \times \left(A_i^{\rm J} J_{ipj} 
    + A_i^{\rm D} D_{ipj}
    + \sum_t A_i^t T_{ip}^t \right),
\end{equation}
and we have multiplied our templates by the \textit{Fermi} exposure, $\mathcal{F}_{ip}$, which describes the angular variation of the sensitivity of \textit{Fermi}.
This step is performed using the Fermi Tools, where we also convolve all templates with the point spread function.
The likelihood of observing $n_{ip}$ counts in pixel $p$ and energy bin $i$ given the mean $\lambda_{ipj}$ is
\begin{equation}
    \label{eq:Poisson likelihood}
    \mathcal{L} \left( n_{ip} | \lambda_{ipj} \right) = \frac{\lambda_{ipj}^{n_{ip}} \exp\left(- \lambda_{ipj}\right)}{n_{ip} ! }.
\end{equation}

As discussed in \cref{sec:DM maps}, we do not know the exact \ac{DM} distribution for a given \csiborg simulation due to the unresolved substructure in halos, although we did not include this uncertainty in \cref{eq:Poisson likelihood}. We model the uncertainty on the substructure contribution to the $J$ factor as a truncated Gaussian.
This choice allows us to marginalise analytically over the substructure uncertainty, such that the conditional probability for a given $\lambda_{ipj}$ given the model parameters is
\begin{multline}
    \mathcal{L} \left( \lambda_{ipj} | A_i^{\rm J}, A_i^{\rm D}, \{A_i^t\}, j \right) = 
    \frac{1}{\sigma_{ipj}} \sqrt{\frac{2}{\pi}} \\ 
     \times \left( 1 + \erf \left( \frac{\mu_{ipj}}{\sigma_{ipj} \sqrt{2}} \right) \right)^{-1}
    \exp \left( - \frac{\left( \lambda_{ipj} - \mu_{ipj} \right)^2}{2 \sigma_{ipj}^2} \right),
\end{multline}
for $\lambda_{ipj} > 0$, and zero otherwise, where
\begin{equation}
    \mu_{ipj} = \mathcal{F}_{ip} \times \left(A_i^{\rm J} \left< J_{ipj} \right> + A_i^{\rm D} \left< D_{ipj} \right> + \sum_t A_i^t T_{ip}^t \right),
\end{equation}
and
\begin{equation}
    \sigma_{ipj} = A_i^{\rm J} \frac{\sigma_{pj}}{J_0}\Delta E_i.
\end{equation}
The likelihood for observing $n_{ip}$ counts in pixel $p$ and energy bin $i$ is then
\begin{multline}
    \label{eq:Hyper likelihood}
    \mathcal{L} \left(n_{ip} | A_i^{\rm J}, A_i^{\rm D},  \{A_i^t\}, j  \right) \\
    = \int \mathcal{L}_{ipj} \left(n_{ip} | \lambda_{ipj} \right) \mathcal{L} \left(\lambda_{ipj} | A_i^{\rm J}, A_i^{\rm D}, \{A_i^t\}, j\right) \dd \lambda_{ipj} \\
     = \sqrt{\frac{2^n}{\pi}} \frac{\sigma_{ipj}^{n_{ip}}}{n_{ip}!} \exp \left(-\frac{\mu_{ipj} ^2}{2 \sigma_{ipj} ^2} \right)\left( 1 + \erf \left( \frac{\mu_{ipj}}{\sigma_{ipj} \sqrt{2}} \right) \right)^{-1} \\
     \times \left(\Gamma
   \left(\frac{n_{ip}+1}{2}\right) \, _1F_1\left(\frac{n_{ip}+1}{2};\frac{1}{2};\frac{\left(\mu_{ipj}
   -\sigma_{ipj} ^2\right)^2}{2 \sigma_{ipj} ^2}\right) \right. \\
    \left. +\sqrt{2} \frac{\left(\mu_{ipj} -\sigma_{ipj} ^2\right)}{\sigma_{ipj}} \Gamma
   \left(\frac{n_{ip}}{2}+1\right) \right. \\
    \left. \qquad \times _1F_1\left(\frac{n_{ip}+2}{2};\frac{3}{2};\frac{\left(\mu_{ipj}
   -\sigma_{ipj} ^2\right)^2}{2 \sigma_{ipj} ^2}\right)\right),
\end{multline}
where $_1F_1$ is the confluent hypergeometric function of the first kind. We describe how we implement this likelihood numerically in \cref{app:Likelihood}.

Assuming that each pixel is independent, the likelihood for the observed data in energy bin $i$, $\mathcal{D}_i$, is
\begin{equation}
     \mathcal{L} \left(\mathcal{D}_i | A_i^{\rm J}, A_i^{\rm D}, \{A_i^t\}, j  \right) = \prod_p  \mathcal{L} \left(n_{ip} | A_i^{\rm J}, A_i^{\rm D},  \{A_i^t\}, j  \right).
\end{equation}

Using the priors, $P$, given in \cref{tab:infered_parameter_summary}, we apply Bayes' identity
\begin{multline}
    \label{eq:Bayes theorem}
        \mathcal{L} \left(A_i^{\rm J}, A_i^{\rm D},  \{A_i^t\}, j| \mathcal{D}_i \right) \\
        = \frac{\mathcal{L} \left(\mathcal{D}_i | A_i^{\rm J}, A_i^{\rm D}, \{A_i^t\}, j \right) P \left( A_i^{\rm J} \right)   P \left( A_i^{\rm D} \right) P\left(\{A_i^t\}\right) P \left(j \right) }{\mathcal{Z}\left(\mathcal{D}_i\right)},
\end{multline}
where
\begin{equation}
    P\left(\{A_i^t\}\right) \equiv \prod_t P\left( A_i^t \right),
\end{equation}
to obtain the posterior, $\mathcal{P} \left(A_i^{\rm J}, A_i^{\rm D}, \{A_i^t\}, j| \mathcal{D}_i \right)$, where $\mathcal{Z}\left(\mathcal{D}_i\right)$ is the evidence and where we consider each energy bin and \csiborg simulation separately. We use the \emcee sampler \citep{emcee} and terminate the chain when the estimate of the autocorrelation length changes by less than 1 percent per iteration and the chain is at least 100 autocorrelation lengths long in all of the parameters.

We apply a Monte Carlo estimate to the likelihood of the \csiborg samples, such that the one-dimensional
posterior for the amplitude $A_i^{\rm X}$ ($X$ being $J$ or $D$) is
\begin{multline}
    \label{eq:spectrum likelihood}
        \mathcal{L} \left(A_i^{\rm X}| \mathcal{D}_i \right) \\
        = \frac{1}{N_{\rm sim}} \sum_j \int \dd \{A_i^t\} \dd A_i^{\rm Y}
        \mathcal{L} \left(A_i^{\rm J}, A_i^{\rm D}, \{A_i^t\}, j| \mathcal{D}_i \right),
\end{multline}
where $Y=J$ if $X=D$ and vice versa.
In practice, we compute the average over \csiborg realisations by first fitting the one-dimensional posteriors $\mathcal{L} \left(A_i^{\rm X}| \mathcal{D}_i, j \right)$ with a spline using the \texttt{GetDist} package \citep{GetDist_2019} and then computing the mean of the resulting functions. This is equivalent to concatenating the Markov Chains if each chain had the same length. 

\subsubsection{Constraining \ac{DM} parameters}

We now have a posterior, $\mathcal{L} \left(A_i^{\rm J}| \mathcal{D}_i \right)$, describing the gamma-ray spectrum from sources that have the same spatial distribution as the $J$ factor. We wish to fit this spectrum to a model, $f_i \left( \bm{\theta} \right)$, for these sources and infer the model parameters $\bm{\theta}$. We assume that our model is deterministic, such that
\begin{equation}
    \mathcal{L} \left(A_i^{\rm J}| \bm{\theta} \right) = \delta \left( A_i^{\rm J} - f_i \left(\bm{\theta}\right) \right),
\end{equation}
and therefore we obtain the likelihood for the observed gamma-ray sky by incorporating \cref{eq:spectrum likelihood},
\begin{align}
        \mathcal{L} \left(\mathcal{D}_i | \bm{\theta}\right) &=
    \int \dd A_i^{\rm J} \, \mathcal{L} \left(\mathcal{D}_i | A_i^{\rm J}\right) \mathcal{L} \left(A_i^{\rm J} | \bm{\theta} \right) \nonumber \\
    &= \int \dd A_i^{\rm J} \, \frac{\mathcal{L} \left(A_i^{\rm J} | \mathcal{D}_i \right) \mathcal{Z}\left(\mathcal{D}_i\right)}{P \left(A_i^{\rm J} \right)} \delta \left( A_i^{\rm J} - f_i \left(\bm{\theta}\right) \right).
\end{align}

We assume that all energy bins are independent such that the likelihood of $\bm{\theta}$ given the full dataset $\mathcal{D}$ is
\begin{equation}
    \mathcal{L} \left(\mathcal{D} | \bm{\theta} \right) = \prod_i \mathcal{L} \left(\mathcal{D}_i | \bm{\theta} \right),
\end{equation}
and so with a final application of Bayes' identity we obtain the posterior for our model parameters
\begin{equation}
    \mathcal{L} \left(\bm{\theta} | \mathcal{D} \right) =
    \frac{\mathcal{L} \left( \mathcal{D} | \bm{\theta} \right)P\left(\bm{\theta}\right)}{\mathcal{Z}\left(\mathcal{D}\right)}.
\end{equation}
If $f_i\left(\bm{\theta}\right)$ comprises exclusively of \ac{DM} annihilation, then, at fixed \ac{DM} mass and annihilation channel, the transformation from the posterior distribution of $A_i^{\rm J}$ to that of $\sigv$ is trivial.
For more complicated models (i.e., where $\bm{\theta}$ consists of more than one parameter), we again calculate the posterior on $\bm{\theta}$ using the \emcee package.

For \ac{DM} annihilation and decay, prompt production, decays, hadronisation and radiative processes associated with the resulting standard model products produce a variety of stable species, including gamma rays. For a given channel, one must know the energy spectrum of the intermediate standard model particles and the resulting branching ratios and energies of the subsequently produced particles. One then has, for each channel, a model for the spectrum of gamma rays as a function of \ac{DM} particle mass and annihilation cross-section or decay rate.

In this work we utilise the pre-computed spectra provided by the \textit{Fermi} collaboration\footnote{\url{https://fermi.gsfc.nasa.gov/ssc/data/analysis/scitools/source_models.html}} which are calculated as described by \citet{Jeltema_2008}. Since we are considering non-relativistic $s$-wave annihilation in this work, one can view the annihilation of two \ac{DM} particles of mass $m_\chi$ as equivalent to the decay of a single particle of mass $2m_\chi$ \cite{PPPC4_2011}. Hence, we obtain the spectrum for decay from the tabulated annihilation spectra by evaluating these at half the relevant particle mass. Furthermore, for kinematic reasons, if we produce two standard model particles, each of rest mass $m_A$, then for decay we enforce the \ac{DM} particle mass to obey $m_\chi > 2 m_A$, whereas for annihilation this is $m_\chi > m_A$.

\begin{table}
    \setlength{\heavyrulewidth}{0.5pt}
    \setlength{\abovetopsep}{4pt}
    \caption{
    Priors on \ac{DM} properties and template amplitudes ($A_i^{t} \in \{A_i^\text{gal}, A_i^\text{iso}, A_i^\text{psc}\}$), as defined in the text. All priors are uniform in the range given, except from the \ac{DM} particle mass, $m_\chi$, since we constrain the cross section, $\sigv$, and decay rate, $\Gamma$, at fixed $m_\chi$. The priors on $A_{i}^{\rm J}$ and $A_{i}^{\rm D}$ depend on the minimum energy of the energy bin, $E_i$, although in all cases the prior is much wider than the posterior. For \ac{DM} decay we also ensure that $m_\chi$ is at least twice the rest mass of the final decay products, and for annihilation this limit is equal to the rest mass of the standard model particle.
    }
    \centering
    \begin{ruledtabular}
        \begin{tabular}{*2c}
            \label{tab:infered_parameter_summary}
            Parameter & Prior \\
            \hline
            $ m_\chi \ / \ {\rm GeV}/c^2 $ & $[2, 500]$\\ 
            $ \sigv \ / \ 10^{-26} {\rm \, cm^3 s^{-1}}$ & $[0, 10^3]$\\ 
            $\Gamma \ / \ 10^{-30} {\rm \, s^{-1}}$ & $[0, 10^3]$ \\
            $A_i^{t}$ & $[0.5,1.5]$ \\
            $A_{i}^{\rm J} \ / \ 10^{-16} {\rm \, cm^{-2} s^{-1} MeV^{-1}}$ & $[0, \left( 300 {\rm \, GeV} / E_i\right)^2]$ \\
            $A_{i}^{\rm D} \ / \ 10^{-16} {\rm \, cm^{-2} s^{-1} MeV^{-1}}$ & [0, $0.5 \times \left( 300 {\rm \, GeV} / E_i\right)^2$] \\
        \end{tabular}
    \end{ruledtabular}
\end{table}

\section{Results}
\label{sec:Results}

\begin{figure}
    \centering
    \includegraphics[width=\columnwidth]{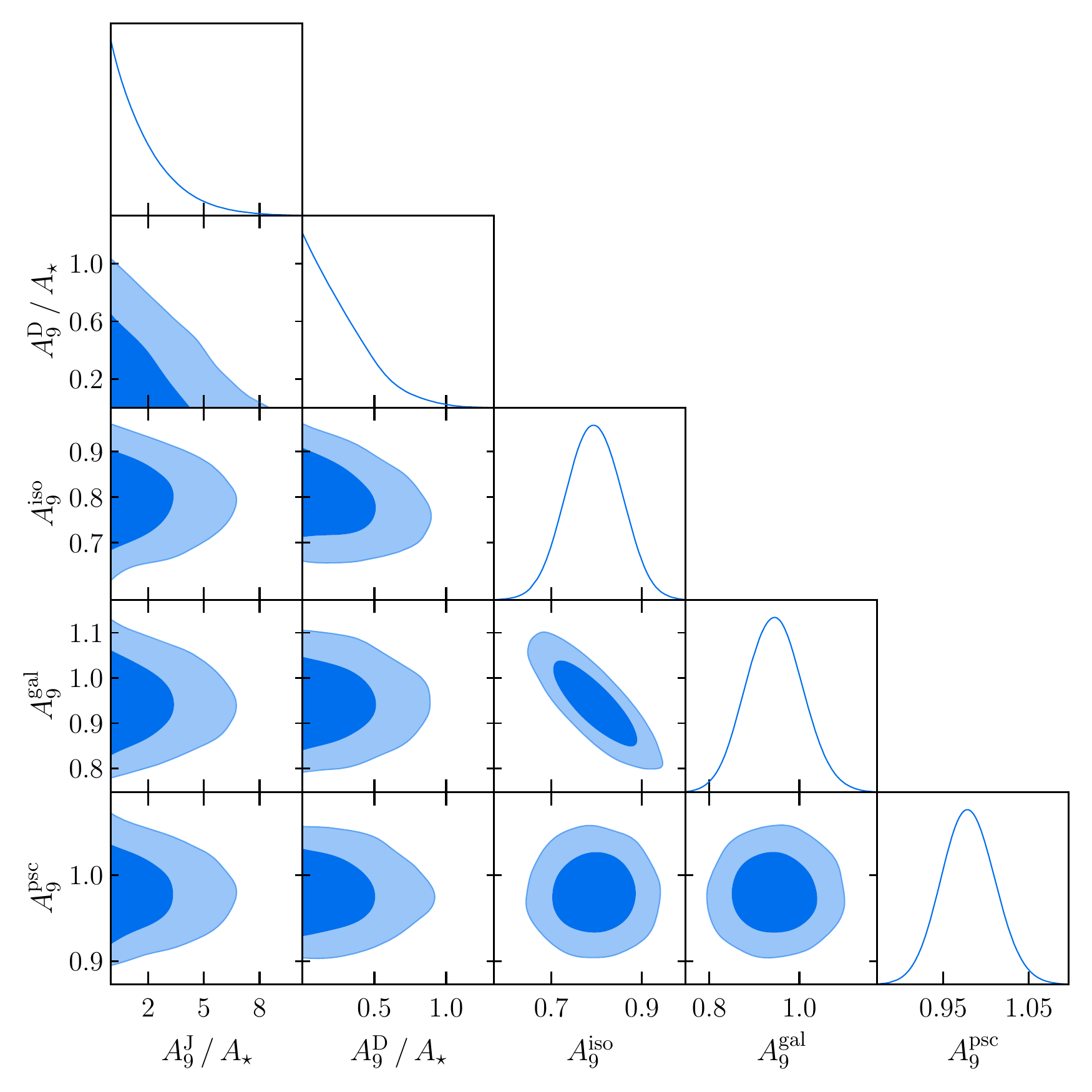}
    \caption{Posterior distributions for \csiborg simulation 7444 of the parameters describing the gamma-ray flux in the energy range $30-50{\rm \, GeV}$. We include templates proportional to the $J$ factor ($A_9^{\rm J}$) and $D$ factor ($A_9^{\rm D})$, as well as isotropic ($A_9^{\rm iso}$), galactic diffuse ($A_9^{\rm gal}$), and point source ($A_9^{\rm psc}$) contributions, and define $A_\star \equiv 10^{-22} {\rm \, cm^{-2} s^{-1} MeV^{-1}}$. The contours show the $1$ and $2\sigma$ confidence intervals.}
    \label{fig:triangle_example}
\end{figure}

\begin{figure}
     \centering
    \subfloat[\empty]{
         \centering
         \includegraphics[width=\columnwidth]{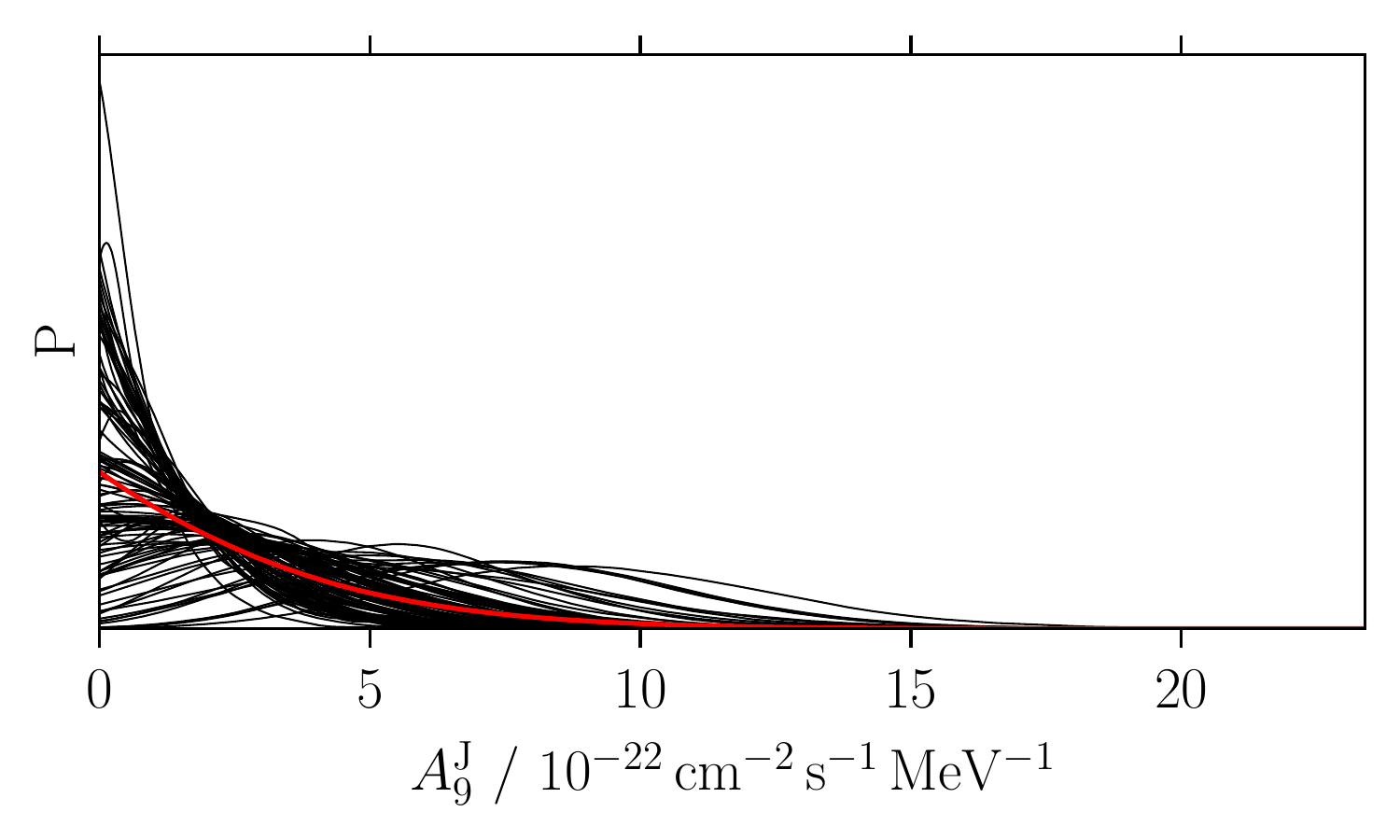}
     }\\\vspace*{-2\baselineskip}
     \subfloat[\empty]{
         \centering
         \includegraphics[width=\columnwidth]{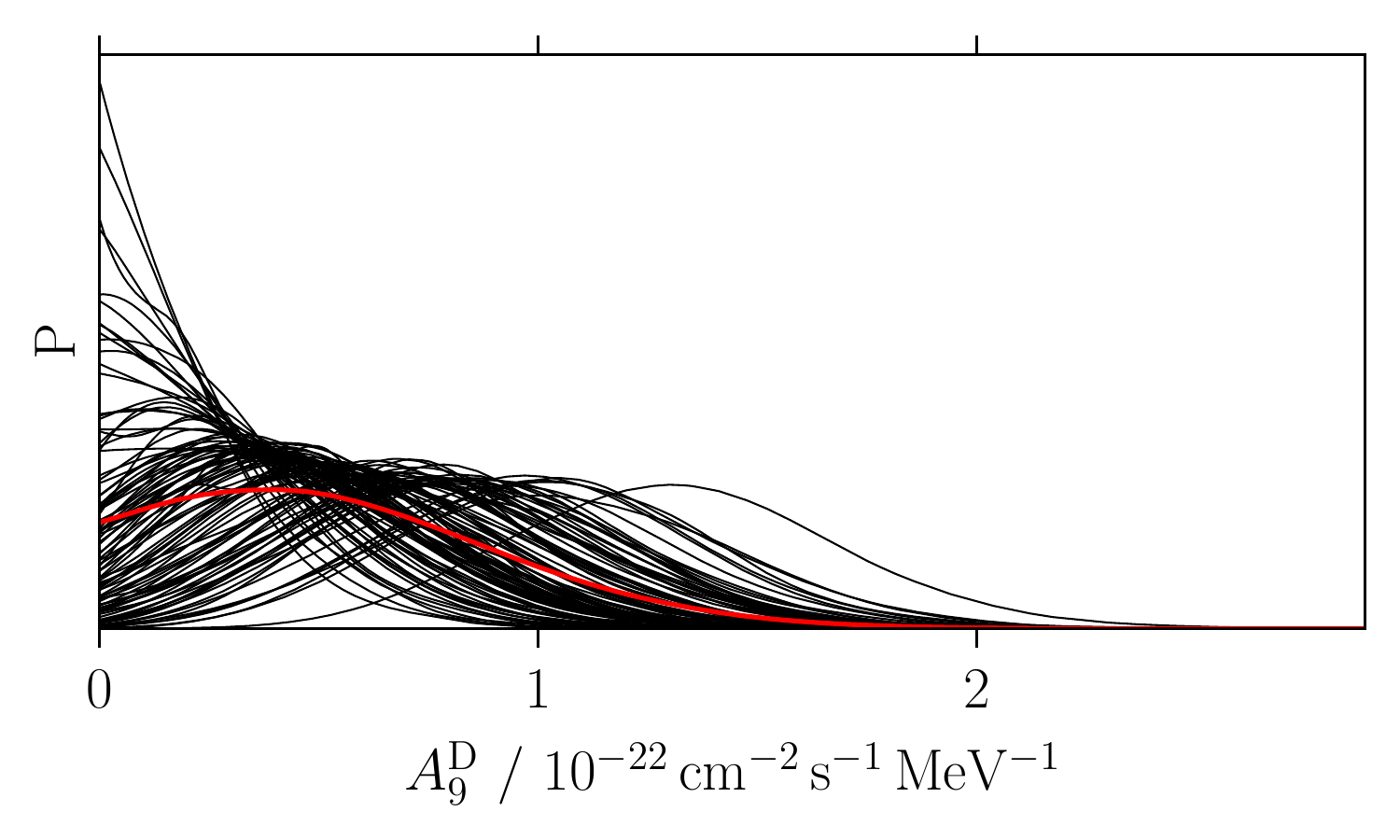}
     }
    \caption{One-dimensional posterior distributions on the coefficients describing the flux proportional to the (upper panel) $J$ factor and (lower panel) $D$ factor in the energy range $30-50{\rm \, GeV}$. Each black line gives the posterior distribution for an individual \csiborg simulation, and the red line is the mean of these, i.e., the final posterior distribution.}
	\label{fig:Combining sims}
\end{figure}

\begin{figure*}
    \centering
    \includegraphics[width=\textwidth]{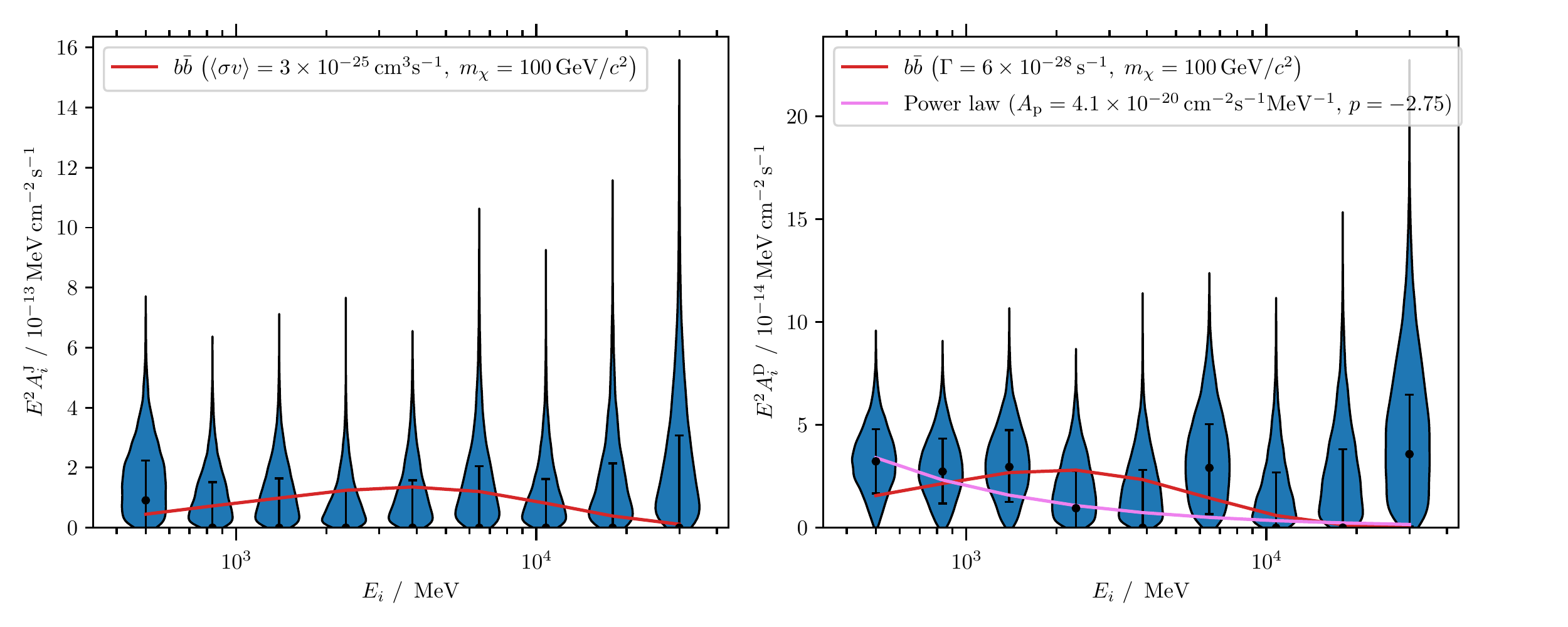}
    \caption{One-dimensional posterior distributions on the coefficients describing the flux in each energy bin, $i$, proportional to (left) the $J$ factor and (right) the $D$ factor. The black points correspond to the maximum likelihood points and the error bars show the $1\sigma$ confidence interval.
    For reference, we plot the expected $A_i^{\rm J}$ and $A_i^{\rm D}$ for \ac{DM} annihilation and decay, respectively, via the $b\bar{b}$ channel for a particle of mass $m_\chi=100{\rm \, GeV}/c^2$ with a thermally averaged cross section of $\sigv = 3 \times 10^{-25} {\rm \, cm^3 s^{-1}}$ and decay rate $\Gamma = 6 \times 10^{-28} {\rm \, s^{-1}}$. We also plot $A_i^{\rm D}$ if the spectrum was due to a power law of amplitude $A_{\rm p} = 4.1 \times 10^{-20} {\rm \, cm^{-2} s^{-1} MeV^{-1}}$ and index $p = -2.75$.
    }
    \label{fig:AJD_violin}
\end{figure*}

\begin{figure}
    \centering
    \includegraphics[width=\columnwidth]{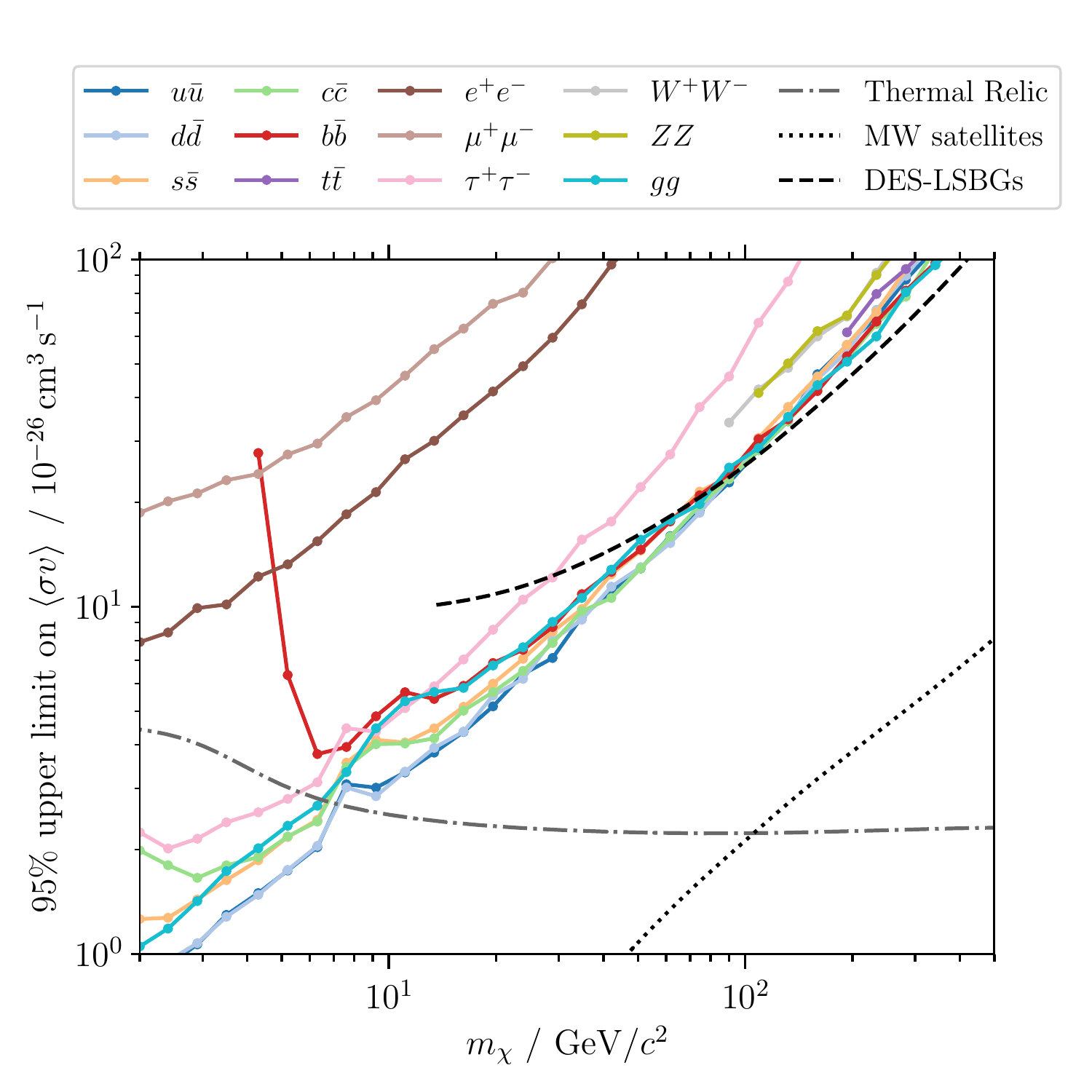}
    \caption{Constraints on \ac{DM} annihilation cross section, $\sigv$, as a function of particle mass, $m_\chi$, for different annihilation channels. The dot-dashed grey line is the expectation for a thermal relic, $\sigv_{\rm th}$, as calculated by \citep{Steigman_2012}. All points below this line rule out the thermal relic cross section at 95\% confidence for the corresponding mass and channel.
    The dotted black line is the constraint obtained by \citet{dSph} from Milky Way satellites for the $b\bar{b}$ channel; we see our constraints are approximately an order of magnitude less stringent. The dashed black line shows the constraints for the $b\bar{b}$ channel derived from the cross-correlation between \textit{Fermi}-LAT and the Dark Energy Survey Y3 low surface brightness galaxy sample (DES-LSBGs) \citep{Hashimoto_2022}. Our field-level inference improves the constraints from large-scale structure by approximately a factor of 2 at $m_\chi = 10 {\rm \, GeV}/c^2$.
    }
    \label{fig:constraint_annih}
\end{figure}

\begin{figure*}
    \centering
    \includegraphics[width=\textwidth]{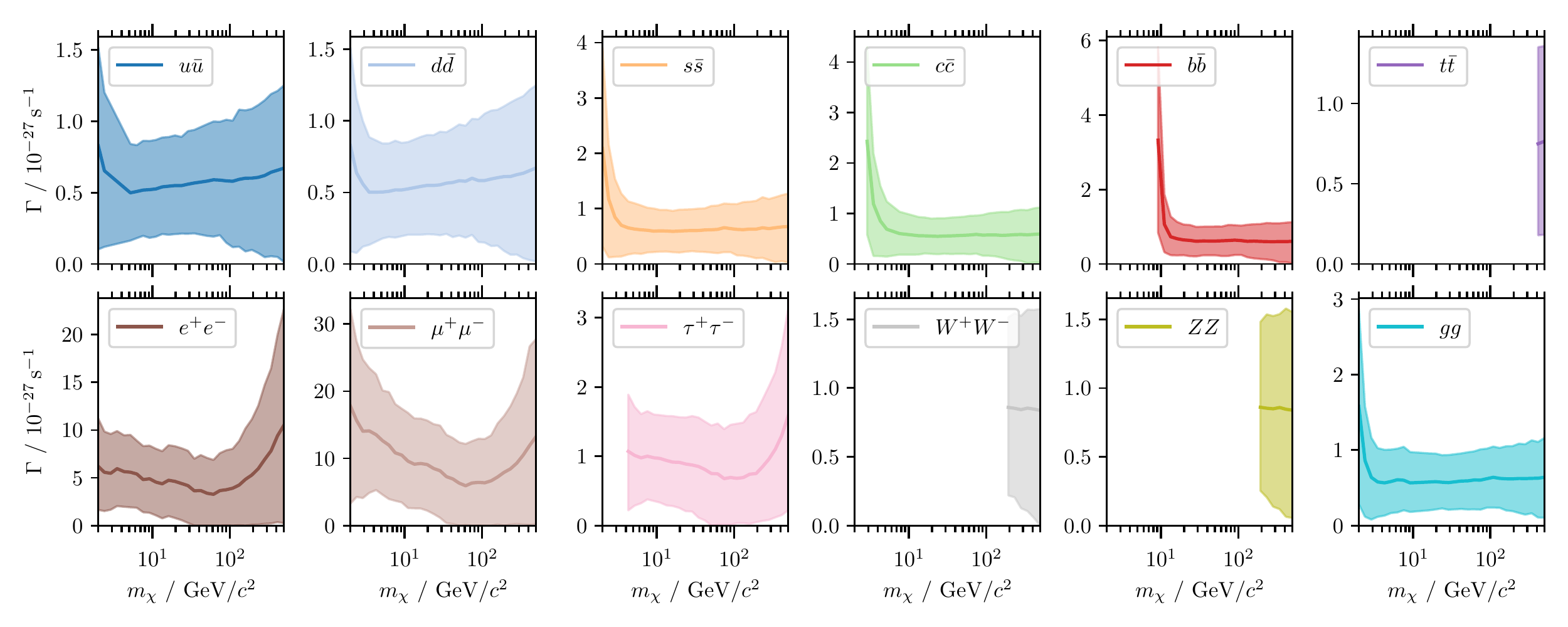}
    \caption{Constraints on \ac{DM} decay rate, $\Gamma$, as a function of particle mass, $m_\chi$, for different decay channels. The solid lines are the median values and the bands show the 95\% confidence regions. We do not infer $m_\chi$, which means that every constraint is conditioned on the corresponding particle mass. For some decay channels and some masses, we see that $\Gamma$ is inferred to be nonzero; however (as shown in \cref{sec:Discussion Decay}), we find that a power-law model better describes the spectrum, suggesting that this flux is not in fact due to \ac{DM} decay. We note that for some channels we cannot probe the full mass range due to the requirement that $m_\chi$ is at least as large as the sum of the masses of the decay products.
    }
    \label{fig:constraint_decay}
\end{figure*}

In \cref{fig:triangle_example}, we show the corner plot for the first stage of our inference, where we infer $A_i^{\rm J}$ and $A_i^{\rm D}$ simultaneously. We emphasise that we fit a different $A_i^{\rm J}$ and $A_i^{\rm D}$ to each energy bin, $i$, and \csiborg simulation. In this example, we consider simulation number 7444 (as given in \citep{max_zenodo}) and the energy range $30-50{\rm \, GeV}$ (energy bin 9).
We see that the parameters corresponding to the isotropic, galactic and point-source contributions are all approximately unity, as one would expect. For this energy bin and \csiborg simulation we see that there is no evidence for a contribution to the gamma-ray flux proportional to either the $J$ or the $D$ factor. We note that $A_i^{\rm J}$ and $A_i^{\rm D}$ are highly degenerate, such that a large value of $A_i^{\rm J}$ corresponds to a small $A_i^{\rm D}$. For our fiducial analysis, we therefore choose to set one of these parameters equal to zero at a time; i.e.\ the inference to place constraints on $\sigv$ will assume $\forall i\; A_i^{\rm D} = 0$ and for $\Gamma$ we assume $\forall i\; A_i^{\rm J} = 0$. This will make our constraints conservative (see \cref{sec:Discussion Annihilation}).

We note that $A_i^\text{iso}$ is strongly degenerate with $A_i^\text{gal}$, which is to be expected since both describe large-scale features across the sky. If we used exactly the same selection criteria as the \textit{Fermi} analysis which produced the non-\ac{DM} templates, then $A_i^\text{gal}$, $A_i^\text{iso}$ and $A_i^\text{psc}$ would all have a mean of unity. This is not true here because the isotropic template is calibrated for latitudes $10^\circ < \left| \lambda \right| < 60^\circ$, whereas we fit our template to $\left| \lambda \right| > 30^\circ$. In general, we find $A_i^\text{iso}$ to be slightly smaller than 1. This is more prominent in the higher energy bins; we find that $A_i^{\rm iso}$ is closer to unity at lower energy. We verify that this is not due to the addition of the $J$ or $D$ factor templates by rerunning the analysis with $A_i^{\rm D} = A_i^{\rm J}=0$ and find that $A_i^{\rm iso}$ remains less than one.

We generate such \ac{MCMC} chains for each of the 101 \csiborg simulations, and we plot the resulting one-dimensional posterior distributions for $A_i^{\rm J}$ and $A_i^{\rm D}$ in \cref{fig:Combining sims}. 
Knowing that each \csiborg simulation is a fair Monte Carlo sample, the final posterior distribution on $A_g^{
\rm X}$ is simply the average of each individual probability distribution, which yields the red lines in the figure.
When marginalised over the \borg chain, we again find $A_i^{\rm J}$ and $A_i^{\rm D}$ are consistent with zero for this energy bin.

This process is repeated for each energy bin to determine the posterior for a given $A_i^{\rm J}$ or $A_i^{\rm D}$, marginalised over all other contributions to the gamma-ray sky and over the uncertainties involved in producing maps of the $J$ and $D$ factors. These spectra are displayed in \cref{fig:AJD_violin}, where we indicate the maximum posterior points by the circles and $1\sigma$ confidence intervals by the error bars. For a given \ac{DM} mass and channel, these posteriors can be trivially transformed into constraints on $\sigv$ or $\Gamma$ for a given energy bin. We then simply multiply the posteriors from each bin to determine our final constraint on these parameters.

In \cref{fig:constraint_annih}, we plot the 95\% upper limit on $\sigv$ as a function of \ac{DM} particle mass, $m_\chi$, for an annihilation which solely produces particle-antiparticle pairs of a single type, but for any standard-model quark, charged lepton or gauge boson (except photons). We compare these constraints to the thermal relic cross section ($\sigv_{\rm th} \approx 3 \times 10^{-26} {\rm \, cm^3 \, s^{-1}}$), such that, if the curve falls below this value in \cref{fig:constraint_annih}, then we rule out \ac{DM} being a thermal relic for the corresponding mass and annihilation channel at 95\% confidence. For all annihilations producing gluons or quarks less massive than the bottom quark, we see that, if \ac{DM} is a thermal relic, it should be more massive than $\sim 7 {\rm \, GeV}/c^2$ since we rule out smaller masses. We are unable to rule out the thermal relic cross section at any mass for production of bottom quarks, top quarks, $W$ bosons or $Z$ bosons. Our constraints for lepton production are much weaker at a given particle mass, such that our constraints for electron or muon production do not cross $\sigv_{\rm th}$. We rule out $\tau$ production for $m_\chi \lesssim 6 {\rm \, GeV}/c^2$ at this cross section.

Turning our attention to \ac{DM} decay, \cref{fig:constraint_decay} shows the inferred decay rate, $\Gamma$, for different decay channels as a function of $m_\chi$. Contrary to our analysis of \ac{DM} annihilation, we find that for the majority of channels we infer a nonzero $\Gamma$ at over $2\sigma$ confidence for at least some $m_\chi$ (corresponding to nonzero $A_i^{\rm D}$ in \cref{fig:AJD_violin}). The results are relatively insensitive to the \ac{DM} particle mass, provided $m_\chi$ is above the threshold for production. For the $b\bar{b}$ channel, we find the inferred $\Gamma$ is $\sim 6 \times 10^{-28} {\rm \, s^{-1}}$, which corresponds to approximately one decay per Hubble time in a volume $\sim 280 {\rm \, km^3}$ at mean cosmological density if $m_\chi = 100 {\rm \, GeV}/c^2$. This is around the smallest $\Gamma$ that has been constrained by any previous study (see \cref{sec:comp_dec}). The inferred $\Gamma$ for decay to the lightest charged leptons is approximately an order of magnitude larger than this.

To determine the overall detection significance, we compute the coefficient for the total flux across all energy bins which multiplies the $D$ factor
\begin{equation}
    A^{\rm D}_{\rm tot} \equiv \sum_i A^{\rm D}_i \Delta E_i = 1.02^{+0.24}_{-0.28} \times 10^{-16} {\rm \, cm^2 s^{-1}}.
\end{equation}
Simply dividing the best fit value by the lower error would suggest that our detection of a contribution to the gamma-ray sky proportional to the $D$ factor has a significance of $3.6\sigma$ when averaged over all available energies. Since our posterior is non-Gaussian, we wish to compute this significance through other methods. We compute the maximum likelihood for each $A_i^{\rm D}$ and, since each energy bin is treated as independent, the maximum likelihood for $A^{\rm D}_{\rm tot}$ is the product of these values. We compare this to the likelihood for $A_i^{\rm D}=0$ and find the change in log likelihood between these two cases is $\Delta \chi^2 \equiv 2 \Delta \ln \hat{\mathcal{L}} = 11.2$, which is equivalent to $3.3\sigma$ for a Gaussian likelihood or a change in the Bayesian Information Criterion (BIC) \citep{Schwarz_1978} of 9.0, if one takes the $A^{\rm D}_{\rm tot} \neq 0$ model as having one more parameter. In \cref{sec:Discussion Decay} we ask whether this is due to \ac{DM} decay, finding that a non-\ac{DM} spectrum is preferred by the data.

\section{Discussion}
\label{sec:Discussion}

In this section, we discuss the possible origin of our results, their limitations and a comparison to existing results in the scientific literature.

\subsection{Interpretation of results}

In this section we investigate which objects and observations drive our results and whether there are non-\ac{DM} explanations for the signal proportional to the $D$ factor.

\subsubsection{Annihilation}
\label{sec:Discussion Annihilation}

 \begin{figure}
     \centering
     \includegraphics[width=\columnwidth]{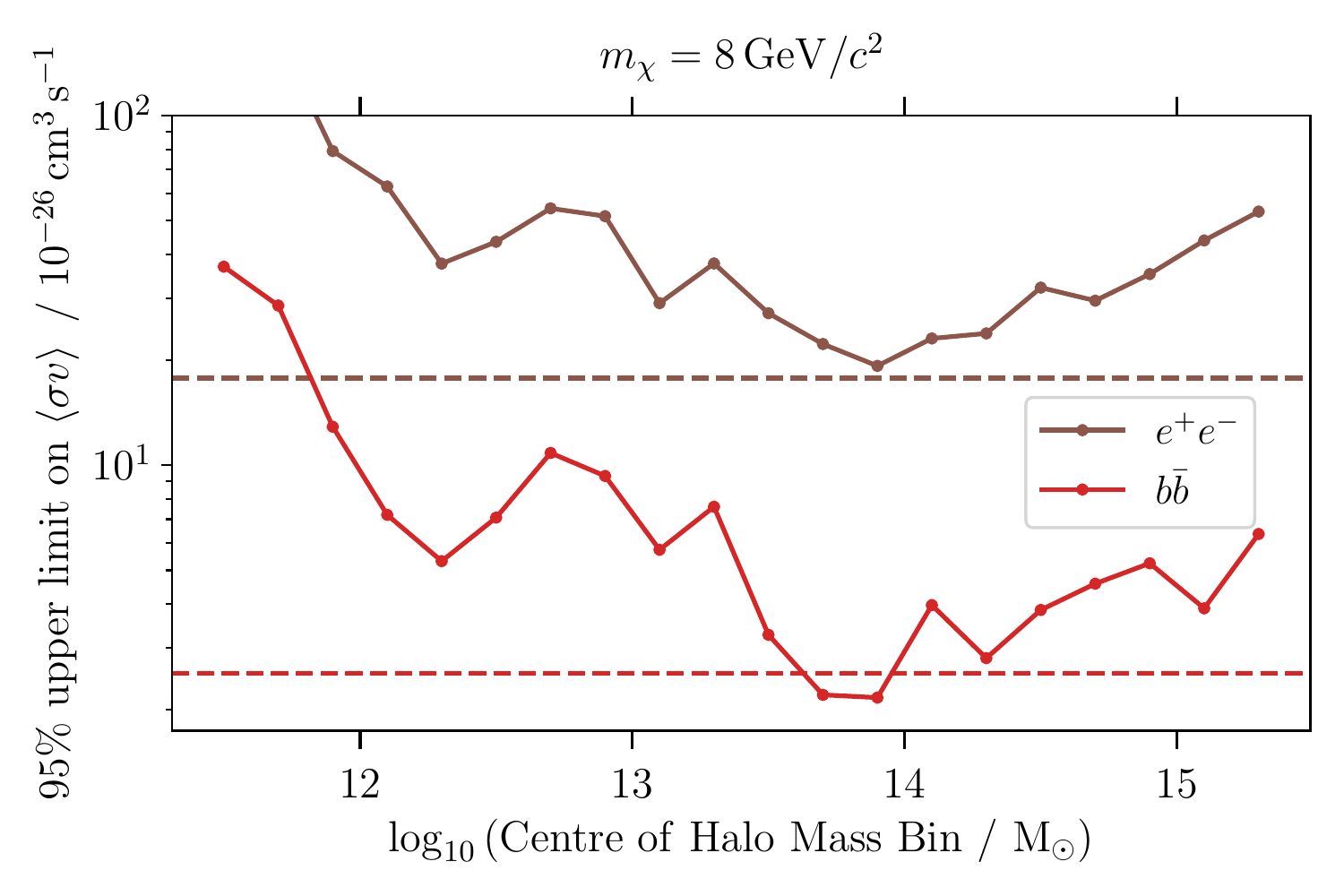}
     \caption{
     Constraints on the $s$-wave self-annihilation cross section, $\sigv$, to either bottom quarks or electrons from halos within a given mass range, where we consider bins in halo mass of width $\Delta \log_{10} \left( M_{\rm h} / {\rm M}_\odot \right) = 1$. Here we use only a single \csiborg simulation (9844); the 95\% upper limits on $\sigv$ for this simulation if we use all halos are indicated by the dashed horizontal lines. Our constraints are dominated by halos of mass $\sim 10^{13.5}-10^{14.5} {\rm \, M}_\odot$.
     }
     \label{fig:mass_binning}
 \end{figure}

 For each channel and $m_\chi$, we compute the change in log likelihood between $\sigv=0$ and the $1\sigma$ constraint on $\sigv$ separately for each energy bin to determine which energy range is dominating our constraint. For \ac{DM} particle masses $m_\chi \lesssim 10 {\rm \, GeV}/c^2$, our constraints are driven by the first three energy bins. For these masses, we expect there to be very few photons produced at high energies at fixed $\sigv$, so these bins are unable to constrain $\sigv$ as very large values are required to produce an appreciable flux. As we move to higher masses, we notice the effect of the data at higher energy, such that at the highest masses we find that the sixth energy bin ($6.5-10.8 {\rm \, GeV}$) is the most constraining. We find a similar trend if we compare using the 95\% confidence limit instead of the $1\sigma$ constraint.

To determine which halos drive our constraints, we produce several $J$-factor maps where each one is only due to objects in a given mass range. We create separate maps for halos in moving bins of width $\Delta \log_{10} \left( M_{\rm h} / {\rm M}_\odot \right) = 1$.  We rerun the inference for a single representative \csiborg simulation (9844) separately for each of these mass bins; i.e., we assume that only a single mass bin contributes to the total $J$ factor. We plot the constraint on $\sigv$ as a function of halo mass in \cref{fig:mass_binning} and observe that the tightest constraints are obtained for halos in the range $\sim 10^{13.5}-10^{14.5}\,\Msun$. If one studied a single object at a fixed distance, then the most massive halo would give the tightest constraints since it has the largest $J$ factor. However, such massive objects are rare, so there is a compromise between having many objects of a given mass across the sky and those objects having a large $J$ factor. Given the tight constraints one can obtain with dwarf galaxies in the Local Group, it is perhaps not surprising that the inclusion of lower mass objects can lead to an improvement in the upper limit on $\sigv$. The inclusion of these structures in this work was possible due to the use of constrained simulations, which provide plausible realisations of these halos given the \ac{ICs} that are constrained on large scales. As a result, \cref{fig:mass_binning} shows the types of objects in the nearby universe one should target to extract maximum information about DM annihilation.
 
Since we found a nonzero flux proportional to the $D$ factor, we rerun the analysis for all \csiborg simulations but infer both $A_i^{\rm D}$ and $A_i^{\rm J}$ simultaneously. In this way, when making our constraint on $\sigv$, we now marginalise over this contribution. Note that in this marginalisation we do not assume a spectral form for the $D$ factor template, such that we marginalise over any source whose spatial distribution is proportional to the local \ac{DM} density, which may or may not be \ac{DM} decay. As anticipated in \cref{sec:Results}, we find that our constraints become tighter, such that our upper limit on $\sigv$ is typically over a factor of 2 smaller. This is as expected: the negative degeneracy between $A_i^{\rm J}$ and $A_i^{\rm D}$ (\cref{fig:triangle_example}) means that, if we allow $A_i^{\rm D}>0$, we must reduce $A_i^{\rm J}$ so that the total flux from these two contributions is approximately constant. When marginalising over $A_i^{\rm D}$, this will result in a posterior on $A_i^{\rm J}$ which is necessarily narrower.
Since these results are tighter than when we set $A_i^{\rm D}=0$, we choose to report the latter as our fiducial results so that our conclusions are conservative.
 
\subsubsection{Decay}
\label{sec:Discussion Decay}

In \cref{fig:constraint_decay} we found a nonzero \ac{DM} decay rate is compatible with the observed gamma-ray sky at over $2\sigma$ confidence for a range of \ac{DM} masses and decay channels. More conservatively, one would say that we find a signal which is proportional to the $D$ factor; i.e., the emitted flux from some region appears to be proportional to the local density, and is compatible with the spectrum of \ac{DM} decay. This source does not necessarily need to be \ac{DM} decay, which can be investigated by fitting the inferred spectrum to a different model. For this we choose a power-law profile, such that the parameter $A^{\rm D}_i$ arises from integrating the spectrum
\begin{equation}
    \frac{\dd N}{\dd E_\gamma} = A_{\rm p} \left( \frac{E_\gamma}{E_0} \right)^p,
\end{equation}
across energy bin $i$, where we normalise to $E_0 \equiv 1 {\rm \, GeV}$. We place broad, uniform priors on $A_{\rm p}$ and $p$ in the range
$[0,10^{-18}] {\rm \, cm^{-2} s^{-1} MeV^{-1}}$
and $[-5,2]$, respectively, and find
 \begin{equation}
 \label{eq:Inferred p}
    \begin{split}
        A_{\rm p} &= \left(4.1 \pm 1.5\right) \times 10^{-20} {\rm \, cm^{-2} s^{-1} MeV^{-1}}, \\
        p &= -2.75^{+0.71}_{-0.46},
    \end{split}
 \end{equation}
 where the limits are at $1\sigma$ confidence. To enable a comparison, we plot the spectrum for this model and a \ac{DM} decay model in \cref{fig:AJD_violin}, where we see that the power law fits better at most energies.
 
To assess the relative goodness of the fit of the two models we compute the BIC. Since we have set deliberately wide priors on our model parameters, ratios of the Bayesian evidence are difficult to interpret. For all channels and masses, we find that the BIC prefers the power-law spectrum, with ${\rm BIC} \geq 1.3$ (the bound is saturated for the gluon channel). Decays via the $t\bar{t}$ channel are least preferred by the data, with ${\rm BIC} \geq 4.0$. We therefore conclude that, although we do find an excess of gamma-ray flux that traces the density of \ac{DM}, its spectrum is fit marginally better by a power law, so this is not evidence for \ac{DM} decay.
 
This conclusion is consistent with previous works studying the origin of the residual gamma-ray flux. The cross-correlation of the gamma-ray sky with galaxy catalogues has been detected at $2-4\sigma$ \citep{Branchini_2017}, and the spectral index of the 1-halo contribution was found to be -2.7 if a single power law is assumed, which is consistent with our inferred index and has a similar significance of detection. Moreover, analysis of the angular power spectrum of the gamma-ray sky \citep{Ackermann_2018} suggests a component that can be modelled as a double power law with an exponential cutoff, with power-law indices $-2.55 \pm 0.23$ and $-1.86 \pm 0.15$, the former of which is consistent with our result. \citet{Ackermann_2018} note that this spectrum is compatible with blazarlike sources being the dominant component at these energies. Since we detect a nonzero contribution with the same spatial variation as our $D$ factor maps, our work suggests that these excesses could be due to sources with a linear bias with respect to the local \ac{DM} density.

\subsection{Systematic uncertainties}
\label{subsec: systematic_uncertainties}

In this section we investigate potential systematic errors in our analysis by changing some of the analysis choices in \cref{sec:Methods}. For computational convenience, throughout this section we use only one \csiborg realisation (simulation 7444 as given in \citep{max_zenodo}) unless otherwise stated.

\subsubsection{Computing the \texorpdfstring{$J$}{J} and \texorpdfstring{$D$}{D} factors}
\label{sec:Sytematics JD maps}

In \cref{sec:Smoothed density field} we computed the $D$ factor and nonhalo contribution to the $J$ factor by smoothing simulation particles onto a grid with a kernel inspired by \ac{SPH}. Besides this kernel, we also consider the \ac{CIC} density assignment in order to quantify the impact of the kernel choice on our constraints. The median change in the constraint on $\sigv$ is 2\% for the $b\bar{b}$ channel if we change to this kernel. For the \ac{DM} decay inference, we find the median and 95\% upper limit on $\Gamma$ change by a median of 16\% and 15\%,
respectively, for this channel. The inferred value of $p$ changes only by $0.01$ when we change to the \ac{CIC} kernel. If the low density regions were driving our constraints, then one would expect large differences between the two procedures, since these regions have the fewest simulation particles and the two kernels have different noise properties for low particle counts. However, we do not see this since the expected flux is highest in the high density regions and the low density regions are relatively unconstraining.

For computational convenience, for our fiducial analysis we chose a \healpix resolution of \nside=256. We rerun the analysis at coarser resolution (\nside=128) and find that our constraints on  $\sigv$ weaken by a median change of 48\% across all masses for the $b\bar{b}$ channel. The median and 95\% upper limit on $\Gamma$ change by a median of 13\% and 10\% respectively for the $b\bar{b}$ channel, and the inferred value of $p$ changes by only 0.03 when we lower the \healpix resolution. It is unsurprising that the $D$ factor analysis is less affected by this choice; for the $J$ factor our template is dominated by high density peaks in the \ac{DM} density field, since the flux is proportional to the square of the density. By using a higher resolution map, one can localise these peaks better to obtain tighter constraints if these are not aligned with peaks in the observed gamma-ray sky.

The $J$ and $D$ factor maps were calculated for each of the 101 \csiborg simulations. By utilising the full suite, we marginalise over the uncertainties in the constrained density modes from both the \borg algorithm and the unconstrained, small-scale modes.
To verify that we have a sufficiently large number of simulations to achieve this, we rerun our analysis 100 times for the $b\bar{b}$ channel using 50 randomly selected simulations to determine a bootstrap uncertainty on our constraints. The standard deviation of the 95\% upper limit on $\sigv$ has a median value of 19\% when considering all masses. The inferred $\Gamma$ has a median bootstrap uncertainty of 4\%. The uncertainty on the inferred power-law index, $p$, is 0.05 and the fractional bootstrap uncertainty on $A_{\rm p}$ is 4\%, which are small compared to the uncertainties we quote in \cref{eq:Inferred p}. We therefore conclude that the number of constrained simulations is adequate.

\subsubsection{Halo density profile}
\label{sec:Systematics Halo Profile}

After identifying halos within the \csiborg simulations, we assumed that all halos have NFW profiles with masses as given by the halo finder and concentrations given by the mass-concentration relation of \citep{Sanchez_Conde_2014}. To determine the sensitivity of our constraints to the assumed profile, we rerun the analysis but assuming that all halos are described by Einasto profiles, as calculated in \cref{sec:Halos}. We find our constraints on $\sigv$ can be up to 80\% tighter if one uses an Einasto profile compared to a NFW. For small mass halos we find that the Einasto profile leads to larger densities near the centre than the NFW profile and, given the importance of these lower-mass objects for our constraint (\cref{sec:Discussion Annihilation}), this leads to smaller values of $\sigv$. Note that a similar tightening of the constraints was observed in \citep{HESS_2} when the Milky Way profile is changed from a NFW to an Einasto profile. We choose to report the most conservative constraints, hence the choice of NFW profiles in our fiducial analysis.

Both the form and parameters of the NFW and Einasto profiles are inspired by $N$-body, \ac{DM}-only simulations. Observations and hydrodynamical simulations suggest that these may not accurately describe the true density profiles. In the presence of baryons, the \ac{DM} profile could be steeper due to adiabatic contraction during galaxy formation \citep{Blumenthal, Gnedin}, or shallower due to the subsequent stellar feedback (e.g. \citep{Pontzen_Governato, DP_CuspCore}). For steeper slopes of the density profile, the $J$ factor near the centre would also be larger, and hence one would expect tighter constraints on $\sigv$. Ideally one would use profiles motivated by hydrodynamical simulations; however, common parameterisations such as \citep{Di_Cintio_2014} apply primarily to sub-Milky Way mass halos while most of those produced in \csiborg are in the group and cluster regimes. Rather than perform a large extrapolation,
we leave it to future work to implement a robust baryonification scheme on these scales.

\subsubsection{Substructure uncertainties}
\label{sec:Systematics substructure uncertainties}

Since the gamma-ray flux from \ac{DM} annihilation is proportional to the square of the density, the substructure of \ac{DM} halos is an important contribution that one must consider; if one computes the angular power spectrum for the $J$ factor, $C_\ell$, one finds that $\left(2 \ell+1\right)C_\ell$ is approximately constant at the smallest scales considered in this work. Usually this substructure is modelled as a mass-dependent multiplicative boost factor \cite[e.g.][]{fornasa2013characterization} and uncertainties captured by looking at the extreme values of the boost for given masses \citep{Fornasa:2016ohl}. We, on the other hand, capture substructure and its uncertainty through \clumpy's probabilistic approach to substructure modelling. This led us to use a non-Poisson likelihood, since we introduced uncertainties, $\sigma_{jp}$, on the Poisson means. To evaluate the impact of this choice, we rerun the analysis assuming a Poisson likelihood by setting $\sigma_{jp} = 0$. We find that our constraints typically change by a few percent across all channels and masses, indicating that the impact of this uncertainty is negligible. However, we note that this could not have been known \textit{a priori}. Although the fractional uncertainties are small near the centres of halos, this is not true in the outskirts, motivating our thorough treatment of uncertainties.

For simplicity, we previously neglected the uncertainty that arises due to stochasticity in the mass-concentration relation. We find our constraints are not very sensitive to the scatter in this relation. For the $b\bar{b}$ channel across all masses, the median change in the constraint on $\sigv$ is 1.4\% if this uncertainty is included.

Another source of systematics due to substructure modelling might be driven by the breakdown of our assumption that the uncertainty in the $J$ factor is Gaussian. Namely, as the considered mass of the clump grows, the total number of such clumps within the host halo decreases. Therefore, it is expected that at some point we transition from the Gaussian into a Poisson regime \citep{lavalle2007clumpiness}. Furthermore, it is not obvious that the contribution to the $J$ factor from these more massive clumps will not outshine the cumulative contribution of the lower-mass clumps. To check for this, we use the \textsc{-h5} module of the \clumpy package to explicitly draw substructure realisations for a typical halo ($M_{\rm h} \approx 5\times 10^{13} \, M_{\odot}$) from the \csiborg simulations. We modify \cref{eqn: mean_J_p} such that
\begin{equation}
    \left< J_p \right> = 
    \langle J_{\mathrm{cont},p}(M_{\rm{th}})\rangle + \langle J_{\mathrm{drawn},p}(M_{\rm{th}})\rangle,
\end{equation}
and
\begin{equation}
    \begin{split}
        \langle J_{\mathrm{cont},p}(M_{\rm{th}})\rangle &= J_{{\rm sm}, p} + \left< J_{{\rm subs}, p}(M_{\rm{th}}) \right> \\
        & \quad + \left< J_{{\rm cross}, p}(M_{\rm{th}}) \right>
        \label{eqn: drawn_subs_threshold},
    \end{split}
\end{equation}
where we introduce
\begin{equation}
    \langle J_{\mathrm{drawn},p}(M_{\rm{th}})\rangle
    =
    \frac{1}{N_{\mathrm{ds}}}
    \sum_{i}^{N_{\mathrm{ds}}} J_{\mathrm{drawn}, pi}(M_{\rm{th}}),
    \numberthis
    \label{eqn: J_drawn}
\end{equation}
with $N_{\mathrm{ds}}$ being the total number of explicit realisations of the clumps with a mass above a given mass threshold, $M_{\rm{th}}$. The quantities $\left< J_{{\rm subs}, p}(M_{\rm{th}}) \right>$ and $\left< J_{{\rm cross}, p}(M_{\rm{th}}) \right>$ from \cref{eqn: drawn_subs_threshold} are obtained by replacing the upper limit of the clump mass distribution by $M_{\rm{th}}$, i.e., replacing $M_2$ with $M_{\rm{th}}$ in \cref{eqn: luminosity_moments}.
To estimate $\langle J_{{\rm drawn} ,p}(M_{\rm{th}})\rangle$, we run $N_{\mathrm{ds}} \approx 1000$ explicit realisations of substructure clumps for a typical halo, requiring that we capture fluctuations in the value of $J_{{\rm sm},p}$ -- the leading contribution to the total $J$ factor of the halo -- at the percent level. In other words, any clump whose contribution to the given pixel will induce a fluctuation to the value of $J_{{\rm sm} ,p}$ of the order of $\sim 1\%$ will be explicitly drawn onto the \healpix grid. This is equivalent to taking one sample from \cref{eqn: clumps_pdf}, but with a modified mass range of the mass function, and setting the lower limit for this draw to be $M_{\rm{th}}$. For more details see section 2.4.3 of \citep{Charbonnier_2012}.

For this experiment, we selected a \healpix resolution of $\nside=1024$, corresponding to the \textit{Fermi}-LAT angular resolution. The corresponding threshold mass for this setup translates to $M_{\rm{th}} = 5.3 \times 10^{9} \, M_{\odot}$ for our chosen halo. We find that
\begin{equation}
    \langle J_{{\rm drawn},p}(M_{\rm{th}})\rangle \sim 0.04 \, \langle J_{\mathrm{cont},p}(M_{\rm{th}})\rangle,
\end{equation}
which justifies our starting assumption of treating the substructure contribution to the total $J$ factor as Gaussian, since the ``drawn'' (Poisson) component is subdominant compared to the ``continuous'' (Gaussian) contribution. 

Note, however, that choosing a smaller $M_{\rm{th}}$, i.e., looking at even smaller fluctuations of $J_{{\rm sm},p}$, would lead to probing the even lower-mass end of the substructure mass function, which would, of course, alter the ratio of the ``drawn'' and ``continuous'' components. However, going below this limit would already enter into a regime where drawing $10^{4}-10^{5}$ clumps from the corresponding version of \cref{eqn: clumps_pdf} will be necessary, which is computationally expensive and well within the validity of the Gaussian approximation. As a comparison, using $M_{\rm{th}} \approx 5.3 \times 10^{9} \, M_{\odot}$ required around $\sim 10^3$ draws. Throughout this section we assumed the same \ac{DM} profile parameterisation and mass-concentration relation as in the fiducial inference (see \cref{sec:Systematics Halo Profile}). The conclusions are unchanged for the same halo using the Einasto profile.

\subsubsection{Non-\ac{DM} templates}
\label{sec:Systematics Non-DM}

The point source template is designed to remove small-scale emission so that large-scale variations in the gamma-ray sky can be modelled more robustly. Since this template is derived from the data, there is a risk that a subset of the point sources could be due to annihilating or decaying regions of high \ac{DM} density. Modelling these as a non-\ac{DM} component would therefore be incorrect. To assess this, we compute the angular cross-correlation function between halos identified from the \csiborg simulations and the positions of point sources detected by \textit{Fermi}-LAT. We find no significant correlation at any scale, justifying our modelling assumptions. This would be expected from the lack of degeneracy between the amplitude of the point-source template, $A_i^{\rm psc}$, and the amplitudes of the $J$ and $D$ factor templates, $A_i^{\rm J}$ and $A_i^{\rm D}$, in \cref{fig:triangle_example}. This suggests that our constraints are not driven by point sources, so the precise model we use for these is not important.

Although in \cref{fig:triangle_example} we see there is little degeneracy between the parameters describing galactic diffuse emission and the \ac{DM} annihilation or decay parameters, one should verify that the results are robust to reasonable variations in these non-\ac{DM} templates. In our fiducial analysis we used the most recent galactic diffuse model provided by the \textit{Fermi} Collaboration (\texttt{gll\_iem\_v07}). We rerun the analysis with an older model (\texttt{gll\_iem\_v02}) and find our constraints on $\sigv$ are slightly weaker, with a median change of 9\% across all mass bins for the $b\bar{b}$ channel, and that the inferred value of $\Gamma$ can vary by $\sim 35\%$. Although there is some variation as we change the model, these are at a similar level to other systematic effects. The inferred power-law index, $p$, only changes by $0.01$ so it is insensitive to this choice.

In our analysis we assume that the amplitudes of the non-\ac{DM} templates are independent, and thus we do not impose a prior on the shape of the spectrum, i.e. on the relative amplitudes between energy bins. If one were to assume that the shape of the spectrum was identical to that in the \textit{Fermi} analysis, then one would enforce the same amplitude for a given template across all bins. If these parameters were strongly degenerate with $A_i^{\rm J}$ and/or $A_i^{\rm D}$, or if their values varied significantly with energy, then imposing such a prior could lead to tighter constraints on $\sigv$ and $\Gamma$. Since neither of these criteria apply (see \cref{fig:triangle_example}), we do not make such a choice, so the spectra of the astrophysical templates are determined empirically. Similarly, we have not imposed the spectra of gamma-rays expected from annihilation or decay, instead deriving constraints from each energy bin separately.

\subsubsection{High redshift sources and optical depth}

\label{subsubsec: Missing_physics}

The gamma rays emitted from either \ac{DM} annihilation or decay would interact with the extragalactic background light (EBL) or \ac{CMB} photons \citep{franceschini2008extragalactic, franceschini2017extragalactic}. This interaction manifests itself through pair production and therefore can cause signal attenuation, which can be modelled through an energy and redshift dependent optical depth coefficient $\tau(E,z)$. Since both the EBL and the \ac{CMB} are approximately isotropic, the optical depth will not have a directional dependence. For the redshift range considered in this paper ($z\lesssim 0.05$), the attenuation of the photon flux due to interaction with background photons will not be significant, except at very high energies ($\sim \mathrm{TeV}$), which lie well above the maximum photon energies we consider here ($\sim 50\, \rm{GeV}$), and thus we neglected this contribution.

Although this is the case for the very nearby Universe, there is also a contribution to $J$ and $D$ from sources outside the \csiborg volume. The expected contribution to the differential photon flux from annihilation is \citep{Slatyer_LesHouches}
\begin{align*}
    \left\langle 
    \frac{\dd^2 \Phi^{\mathrm{ann}}}{\dd E_\gamma \dd \Omega} 
    \right\rangle &= 
    \frac{\langle \sigma v \rangle \bar{\rho}^2_{\mathrm{DM},0}}{8 \pi m_{\chi}^2}
    \int \dd z 
    \left.\left(\frac{\dd N_{\gamma}}{\dd E^\prime_{\gamma}}
    \right)\right|_{E^\prime_{\gamma} = E_\gamma(1+z)}\\
    &\times \frac{(1+z)^3}{H(z)}
    e^{-\tau(E^\prime_{\gamma},z)}
    \langle 
    \left( 1 + \delta (z, \Omega) \right)^2 
    \rangle,
    \numberthis
    \label{eqn: diff_flux_with_EBL_abs}
\end{align*}
where $\bar{\rho}_{\mathrm{DM},0}$ represents present-day \ac{DM} density and $\delta(z,\Omega)$ is the density fluctuation. This can be directly computed from the nonlinear matter power-spectrum (see, for example, \citep{serpico2012extragalactic}) or by using the halo model approach \citep{Ando:2005xg, Hutten_2018}. Within the halo model, this factor comes directly from averaging the one-halo annihilation luminosity over the halo mass function. This is equivalent to marginalising over plausible realisations of the \ac{DM} distribution in our Universe by utilising the Press-Schechter \citep{Press_Schechter74} formalism, or any other halo-formation model.

As in \citep{Hutten_2018} (see Fig.~10 in their Appendix B), we estimate the integrand at a given $E_\gamma$ and integrate between $z=0.05$ and $z=10$ to determine the ratio of this contribution to that explicitly modelled from \csiborg. We find that this ratio is approximately unity at $E_\gamma = 5 {\rm \, GeV}$ for $m_\chi = 10{\rm \, GeV}/c^2$ for the $b\bar{b}$ channel. One may be concerned that this is an important contribution; however, since our chosen \healpix resolution of $\nside=256$ corresponds to a physical scale of $\sim 0.6 \Mpch$ at the edge of the \csiborg volume, one would expect that the extragalactic sources beyond $z \gtrsim 0.05$ are unresolved, and therefore this contribution will almost entirely be absorbed into the isotropic template.
Of course, clustering of sources at redshifts beyond the \csiborg volume would lead to an anisotropy in this unresolved emission. Our constraints are completely independent of how one models the isotropic part of the high-redshift component, and we leave it to further work to model the fluctuations about this, for example, by including constrained realisations of the density field for larger volumes. Since this contribution can only increase the $J$ factor, we always underestimate the $J$ factor in our templates, making our constraints conservative.

We note that in \cref{eqn: diff_flux_with_EBL_abs} one must correct for the redshift of emission; i.e., the spectrum should be evaluated at $E_\gamma \left( 1 + z \right)$ for a source at redshift $z$ if we observe a photon at energy $E_\gamma$. Since we only considered sources at $z \lesssim 0.05$, we neglected this effect. If we consider the extreme case where all our sources were actually at $z=0.05$, we would find that our estimate of the flux at a given $\sigv$ is correct to within 5\% in the five lowest energy bins for the $b\bar{b}$ channel at $m_\chi = 100 {\rm \, GeV}/c^2$ (similar effects are seen for other channels and masses). Since this is comparable to the size of other reasonable variations to the model and this is an unrealistically extreme case, we are justified in making this assumption.

\subsection{Comparison to literature}
\label{sec:comp}

To enable a comparison between the constraints on $\sigv$ and $\Gamma$ obtained in this work using large-scale structure and those from the literature, we now briefly summarise other methods for inferring these parameters and the results they produce.

\subsubsection{Annihilation}
\label{sec:comp_anh}

The release of energy by annihilating \ac{DM} has the potential to affect several observables over the Universe's history. The earliest important observable is the element yield from \ac{BBN}: a $100 {\rm \, GeV}/c^2$ \ac{DM} particle with the thermal relic cross section would release $\sim$1 MeV of energy for every baryon in the Universe per Hubble time during the \ac{BBN} era. This has the potential to alter subdominant nuclear reactions, although the effect is not strong enough to lead to stringent constraints \citep{BBN_rev, BBN}. The next important epoch is recombination, where annihilating \ac{DM} has the potential to ionise a non-negligible fraction of the hydrogen in the Universe. This would absorb \ac{CMB} photons after recombination, to which the \ac{CMB} angular power spectra and power spectra are acutely sensitive \citep{CMB}. This allows thermal relic \ac{DM} with a velocity-independent cross section to be ruled out for masses below $\sim 10-30 {\rm \, GeV}/c^2$ depending on the annihilation channel \citep{Planck, Planck_Slatyer}. Of course, these bounds could be evaded by a large branching fraction into neutrinos or other particles with no electromagnetic interaction. The \ac{CMB} constraints are particularly important for light \ac{DM} ($m_\chi \lesssim 1 {\rm \, GeV}/c^2$) where the effective area and angular resolution of telescopes such as \textit{Fermi}-LAT are poor. At even later times ($2 \lesssim z \lesssim 6$), observations of the Lyman-$\alpha$ forest constrain the gas temperature (e.g., \citep{Lya_1, Lya_2}), which would be increased by annihilations, although this has not been used to set quantitative constraints.

These bounds were derived purely by considering the effects of energy injection into the Universe, but more information is available from observations of the potential annihilation products themselves. This is most often done by means of high-energy photons (a common final product regardless of annihilation channel), and forms the context for our own analysis. Of course, the spectra of the final-state photons depend crucially on the channel, as we have described previously. In the local Universe, the most promising targets are the galactic centre, nearby groups or clusters, and dwarf galaxies in the Local Group. The former is the greatest nearby concentration of \ac{DM}, but also suffers from large astrophysical backgrounds, and the expected signal depends sensitively on the poorly known \ac{DM} density profile of the Milky Way. Nevertheless, there are claims for a gamma-ray excess that could be due to annihilating \ac{DM} \citep{Hooper_2011}. In particular, \citet{Hooper_2011} claim the excess is well fit for $m_\chi \sim 7-10 {\rm \, GeV}/c^2$ with $\sigv \sim \left(0.5-5 \right) \times 10^{-26} {\rm \, cm^3s^{-1}}$ annihilating via the $\tau^+\tau^-$ channel. Although we cannot rule out the lower values of $\sigv$ for this mass range, we do find that $\sigv < 4.5 \times 10^{-26} {\rm \, cm^3s^{-1}}$ at 95\% confidence for these masses and this channel, which is incompatible with the larger values of $\sigv$ reported. Clusters are also massive accumulations of \ac{DM} and permit a statistical analysis, but also suffer from potentially significant backgrounds. Dwarf galaxies, although smaller and less dense, have a much lower baryonic mass and hence the lowest contribution from degenerate astrophysical effects, affording a cleaner test. However, one is limited to a small sample size and thus one has to assume that the objects do not have peculiarities, e.g. unusual boost factors. By looking at a larger number of sources, as we do here, one can average over a more representative sample of substructure.

\textit{Fermi}-LAT has been used to set limits on the annihilation cross section using dwarf galaxies, the Milky Way halo \citep{Springel_2008, MW_halo_0, MW_halo}, and galaxy groups \citep{groups}. The strongest constraints come from the dwarfs, which, due to their lower distances, offer higher peak signals than clusters \citep{dwarf_vs_cluster}. These have been used to rule out the thermal relic cross section for masses below $\sim 100 {\rm \, GeV}/c^2$ assuming annihilation to $b$ quarks \citep{dSph_0, dSph}, as depicted in \cref{fig:constraint_annih} (although see \citep{Ando_2020}). Even stronger constraints, ruling out the thermal relic scenario to $\mathcal{O}($TeV) mass scales for annihilation to $b\bar{b}$, have been claimed from a radio search of the Large Magellanic Cloud \citep{radio}. Somewhat weaker constraints have also been obtained using dwarf irregular galaxies \citep{dIrr} and by cross-correlating \textit{Fermi}-LAT data with the positions of nearby galaxies without knowledge of those galaxies' distances (dashed line in \cref{fig:constraint_annih}) \citep{Hashimoto_2021,Hashimoto_2022}. Further information can be gleaned by cross-correlating gamma-ray flux with a tracer of density such as gravitational lensing \citep{Ammazzalorso_2020}.

Data from the ground-based air Cherenkov telescopes VERITAS, MAGIC, HAWC, and H.E.S.S. have also been used to set constraints from dwarfs, which dominate those from \textit{Fermi}-LAT for $m_\chi \gg 1 {\rm \, TeV}/c^2$ \citep{MAGIC, VERITAS, HAWC, HESS}. H.E.S.S. has also been applied to the galactic centre, achieving stronger constraints at very high energies at the cost of increased systematic uncertainty due to astrophysical backgrounds \citep{HESS_1, HESS_2}. It is worth noting also that constraints on both annihilation and decay can be set by direct detection laboratory experiments, although these are considerably weaker than astrophysical constraints \citep{Marrodan_Undagoitia_2021}.

Annihilating \ac{DM} produces other cosmic rays besides photons, most notably positrons and antiprotons. The AMS-02 instrument has provided data on the spectrum of a wide range of cosmic ray species \citep{AMS_positron, AMS_antiproton, AMS_antiproton_2}. Despite uncertainties due to cosmic ray propagation and the impact of the Sun's magnetic field, antiproton observations have been used to set bounds that beat those from \textit{Fermi}-LAT in some cases, for example, constraining the $\mu^+\mu^-$ channel to $m_\chi\sim100 {\rm \, GeV}/c^2$ at the thermal relic cross section \citep{AMS_antiproton}. It is also possible to search for gamma-ray lines, which are generically expected to be weak but may be prominent if the \ac{DM} particle decays to charged particles similar to it in mass. Line limits from \textit{Fermi}-LAT and H.E.S.S. are presented in \citep{HESS_2, HESS_line_2} and \citep{Fermi_line_1, Fermi_line_2} respectively.

\subsubsection{Decay}
\label{sec:comp_dec}

Similar considerations to those of Sec.~\ref{sec:comp_anh} allow cosmological constraints to be placed on \ac{DM} decay. These constraints are stronger at lower redshifts, where a greater fraction of \ac{DM} decays per Hubble time for fixed decay rate. This allows \ac{BBN} to test decay lifetimes around $10^{18}$s, the \ac{CMB} $10^{25}$s, and the Lyman-$\alpha$ forest $10^{25}-10^{26}$s \citep{Slatyer_LesHouches}.

Constraints can also be derived from gamma-ray and neutrino telescopes. In particular, data from \textit{Fermi}-LAT, AMS-02, PAO, KASCADE, and CASAMIA have been used to constrain the \ac{DM} lifetime at the $10^{27}-10^{28}$s level for $10^2{\rm \, GeV}/c^2 < m_\chi < 10^{17} {\rm \, GeV}/c^2$ \citep{decay_1, decay_2}.
For lower-mass \ac{DM} decaying primarily leptonically, bounds at the $10^{25}-10^{26}$s level can be set from X-ray and gamma-ray telescopes, the spectrometer on board the Voyager I spacecraft, and the heating of gas-rich dwarf galaxies, as well as the Lyman-$\alpha$ forest and \ac{CMB} as described above \citep{decay_3, dwarf_heating}. These constraints imply that over a very large \ac{DM} mass range, only a tiny fraction of the total \ac{DM} can decay during the lifetime of the Universe. Decaying \ac{DM} can also be constrained using the masses and abundances of Milky Way satellites in case the decay gives momentum to the \ac{DM} particle, which provides a constraint of order the age of the Universe (e.g., \citep{Nadler}). The inferred values of $\Gamma$ in this work are compatible with these constraints.

\subsection{Future directions}

\subsubsection{Including additional mass}

Our analysis deliberately targets large-scale structure as a source of annihilation or decay flux in order to be fully complementary to studies of particular objects while avoiding their systematics. This has made our constraints conservative because significant $J$ and $D$ factor contributions come from the Milky Way halo and dwarf spheroidals in the Local Group. Incorporating these into our mass model would therefore produce the most constraining results possible, modelling flux from all mass in the local Universe. For the Milky Way this could be done by detailed modelling of the properties of our host halo along the lines of \citep{MW_halo_0, MW_halo} but using the latest data from \textit{Gaia} \citep{Nitschai_2020,Nitschai_2021}; this will be the subject of future work. A separate likelihood component could be added for local dwarf galaxies (cf. \citep{dSph_0, dSph}). We note that inferring \ac{ICs} which could produce such structures with the correct masses and locations using a process similar to the \textsc{sibelius} simulations \citep{Sawala_2022,McAlpine_2022} would be a computationally demanding task. A more feasible approach may be to populate larger halos with such objects \textit{a posteriori} in a manner similar to how one paints galaxies onto a $N$-body simulation.

\subsubsection{Velocity dependence}

In this work we have assumed that $\sigma v$ is independent of energy. One can generalise this such that the cross section is multiplied by a function $S$ of the relative velocity between two \ac{DM} particles, $v_{\rm r}$, i.e. $\left( \sigma v \right) = \left( \sigma v \right)_0 S \left(v_{\rm r} / c \right)$. The velocity-dependent term is commonly modelled as $S(x)=x^n$, where in this work we have considered $n=0$ ($s$-wave) scattering. Other velocity dependencies are theoretically interesting: in models with minimal flavour violation, $n=2$ ($p$-wave) annihilation dominates for Majorana fermions forming Standard Model fermion-antifermion pairs, since the $s$-wave is chirality suppressed \citep{Kumar_2013}. A null result for $p$-wave annihilation in the galactic centre is presented in \citep{Johnson_2019}. Similarly, $n=4$ ($d$-wave) dominates in such models if \ac{DM} is instead a real scalar \citep{Giacchino_2013,Toma_2013}. Because of the small \ac{DM} velocities within halos, one would expect these signals to be harder to detect than $s$-wave scattering. However, the larger velocity dispersion in massive objects increases the relative importance of high mass objects relative to local ones \citep{Baxter_2022}, making extragalactic halos interesting targets to probe velocity-dependent \ac{DM} annihilation. We will analyse these models using our constrained simulations in a future publication.

Furthermore, if \ac{DM} has long-range self-interactions, then the annihilation is Sommerfeld enhanced \citep{Sommerfeld_1931}, corresponding to $n=-1$ and thus a high, potentially detectable annihilation rate.

When studying these velocity-dependent cross sections, one cannot use the $J$ factor given in \cref{eq:J def}, but instead \citep{Boddy_2019}
\begin{equation}
    J_S = \int \dd s \, \dd^3v_1 \, \dd^3v_2 \,
        S \left( \frac{\left| \bm{v}_1 - \bm{v}_2 \right|}{c} \right) 
        f \left( \bm{r}, \bm{v_1} \right)
        f \left( \bm{r}, \bm{v_2} \right),
\end{equation}
where $f\left( \bm{r},\bm{v} \right)$ is the distribution function of \ac{DM} particles. This thus requires one to know or model the velocities of \ac{DM} particles within halos and is therefore left to future work, although we note that $J_S$ has recently been calculated for a range of \ac{DM} density profiles \citep{Boucher_2021}.

\section{Conclusions}
\label{sec:Conclusions}

Indirect detection of \ac{DM} annihilation or decay through gamma-ray emission has previously typically involved inference from a small number of nearby, \ac{DM} rich objects (the Milky Way, dwarf spheroidals in the Local Group or local groups and clusters) or by cross-correlating the gamma-ray background with other catalogues. Instead, in this work we utilise the \csiborg suite of constrained simulations of the local $155\Mpch$ to forward model the predicted gamma-ray sky for $s$-wave \ac{DM} annihilation or decay due to the large-scale structure. We marginalise over uncertainties in the density field reconstruction, unresolved substructure, and non-\ac{DM} contributions to the signal, and compare to data from \textit{Fermi}-LAT via a \ac{MCMC} algorithm.

We rule out the thermal relic cross section at 95\% confidence for \ac{DM} particles of mass $m_\chi \lesssim 7 {\rm \, GeV}/c^2$ whose annihilation produces gluons or quarks less massive than the bottom quark. Our constraints for the production of charged leptons are approximately an order of magnitude less stringent, and we are unable to rule out the thermal relic cross section for the production of top or bottom quarks in our fiducial analysis. We infer at $3.3\sigma$ a nonzero contribution to the gamma-ray sky with the same spatial distribution as predicted by \ac{DM} decay. For the decay to quarks, this corresponds to a decay rate of $\Gamma \approx 6 \times 10^{-28} {\rm \, s^{-1}}$. However, we find that a power-law spectrum is preferred by the data,
and we infer that the power-law index is $p=-2.75^{+0.71}_{-0.46}$. If we marginalise over the contribution with the same spatial distribution as \ac{DM} decay, we obtain constraints on $\sigv$ which are twice as tight as our fiducial analysis.

Our constraints on the annihilation cross section are less stringent than those obtained by studying other objects, such as the GCE or dwarf galaxies in the Local Group. Given the sensitivity of the dwarf spheroidal analysis to the prior on galaxy mass \citep{Ando_2020} and the conflicting explanations for the GCE, this work provides a useful independent probe of novel \ac{DM} properties by forward modelling the whole gamma ray sky and will thus be sensitive to different systematics. The field-based framework we develop implicitly incorporates not just the two point correlation function -- a more traditional way to constrain \ac{DM} properties from large-scale structure -- but all other higher order statistics as well. Since both \ac{DM} annihilation and decay fluxes are determined by line-of-sight integrals of the density field, the use of constrained simulations provides a convenient way of calculating these integrals for the observed Universe. Future work should be dedicated to a joint inference where one combines the contribution to the \ac{DM} annihilation or decay signal from large-scale structure with objects from the Local Group. As with analyses on smaller scales, there is some sensitivity to how one parametrises the halo density profiles, and thus future analysis should include procedures to (probabilistically) model baryonic effects on the \ac{DM} density profile.

\section*{Data availability}

This work required modifications to the \clumpy package in order to model substructure in extragalactic halos and process all the halos from \csiborg at once. These modifications will be included in the next \clumpy release. Our $J$ and $D$ factor maps may be useful for other indirect detection analyses of current or future data, so we make these maps publicly available at \url{https://cloud.aquila-consortium.org/s/csiborg_jdmaps}.

\acknowledgements
{
We thank David Alonso, Eric Baxter, C\'{e}line Combet, Sten Delos, Julien Devriendt, Seth Digel, Scott Dodelson, Pedro Ferreira, Moritz H\"{u}tten, Harley Katz, Eiichiro Komatsu, David Maurin, Samuel McDermott, Lance Miller, John Peacock, Martin Rey, Daniela Saadeh and Fabian Schmidt for useful input and discussions. We thank Jonathan Patterson for smoothly running the Glamdring Cluster hosted by the University of Oxford, where most of the data processing was performed.

D.J.B. is supported by STFC and Oriel College, Oxford. A.K. acknowledges support from the Starting Grant (ERC-2015-STG 678652) ``GrInflaGal'' of the European Research Council at MPA. H.D. was supported by St John's College, Oxford, a McWilliams Fellowship at Carnegie Mellon University, and a Royal Society University Research Fellowship (Grant No. 211046).
J.J. acknowledges support by the Swedish Research Council (VR) under the project 2020-05143 -- ``Deciphering the Dynamics of Cosmic Structure". G.L. acknowledges support by the ANR BIG4 project, grant ANR-16-CE23-0002 of the French Agence Nationale de la Recherche. D.J.B., G.L. and J.J. acknowledge  support by the Simons Collaboration on ``Learning the Universe''.
This project has received funding from the European Research Council (ERC) under the European Union’s Horizon 2020 research and innovation programme (Grant Agreement No. 693024).
This work was done within the Aquila Consortium (\url{https://www.aquila-consortium.org/}).

Some of the results in this paper have been derived using the
\healpy and \healpix \citep{Zonca_2019,Gorski_2005},
\clumpy \citep{Charbonnier_2012,Bonnivard_2016,Hutten_2019}, 
\emcee \citep{emcee}, and
\fermipy \citep{FermiPy}
packages.
}

Supporting research data are available on reasonable request from the corresponding authors.

For the purpose of open access, the authors have applied a Creative Commons Attribution (CC BY) licence to any Author Accepted Manuscript version arising.

\bibliographystyle{apsrev4-1}
\bibliography{references}

\appendix

\section{Numerical Implementation of Likelihood}
\label{app:Likelihood}

When evaluating the likelihood in \cref{eq:Hyper likelihood}, we often care about small fractional errors on the model, such that the parameter
\begin{equation}
    x \equiv \frac{\left(\mu - \sigma^2 \right)^2}{2 \sigma^2}
\end{equation}
is large, where we have omitted indices for clarity. We therefore must evaluate $_1F_1(a,c,x)$ for $x \gg 1$, although this is problematic since $_1F_1$ has the asymptotic expansion \citep{AbramowitzStegun_1972}
\begin{equation}
    _1F_1 \left (a,c,x \right) \sim \frac{\Gamma \left( c \right)}{\Gamma \left( a \right)} e^{x} x^{a-c},
\end{equation}
and thus diverges as $x\to \infty$. We therefore define the function
\begin{equation}
    f \left( a, c, x \right) \equiv e^{-x} x^{c-a} {_1F_1} \left (a,c,x \right),
\end{equation}
which is finite as $x \to \infty$. Thus the likelihood for a single pixel becomes
\begin{equation}
    \label{eq:alternative likelihood}
    \begin{split}
        \mathcal{L} \left( n \right)
    & = \frac{\left( \mu - \sigma^2 \right)^n}{\sqrt{\pi}n!}
    \exp \left( \frac{\sigma^2}{2} - \mu \right)
    \left( 1 + \erf \left( \frac{\mu}{\sqrt{2}\sigma} \right) \right)^{-1} \\
    & \times \left( \Gamma\left( \frac{1+n}{2} \right) g_n \left( x \right)  + 2 \Gamma\left( 1 + \frac{n}{2} \right) h_n \left( x \right) \right),
    \end{split}
\end{equation}
where we have defined the functions
\begin{equation}
    g_n \left( x \right) \equiv f \left( \frac{1+n}{2}, \frac{1}{2}, x \right),
\end{equation}
and
\begin{equation}
    h_n \left( x \right) \equiv  f \left( 1 + \frac{n}{2}, \frac{3}{2}, x \right).
\end{equation}
Given the asymptotic expansion of $_1F_1$, it is clear that \cref{eq:alternative likelihood} is equivalent to the Poisson distribution in the limit $\sigma \to 0$ at fixed $\mu$.

If $n$ is even, then $g_n (x)$ can be expressed as
\begin{equation}
    g_n \left( x \right) = \sum_{m=0}^{n/2} \alpha_m^{\left( n \right)} x^{-m},
\end{equation}
where the coefficients obey the recurrence relation
\begin{equation}
    \alpha_m^{\left( n + 2 \right)} = \frac{2}{n+1} \left( \left( n - m + \frac{3}{2}\right) \alpha_{m-1}^{\left( n \right)}  + \alpha_m^{\left( n \right)} \right),
\end{equation}
and
\begin{equation}
    \alpha_0^{\left( 0 \right)} = 1.
\end{equation}
Similarly, if $n$ is odd, then
\begin{equation}
    g_n \left( x \right) = \sum_{m=0}^{(n-1)/2} \beta_m^{\left( n \right)} x^{-m},
\end{equation}
for
\begin{equation}
    \beta_m^{\left( n + 2 \right)} = \frac{2}{n+2} \left(  \left( n - m + \frac{3}{2} \right) \beta_{m-1}^{\left( n \right)} + \beta_m^{\left( n \right)} \right),
\end{equation}
and
\begin{equation}
    \beta_0^{\left( 1 \right)} = 1.
\end{equation}
For odd $n$, the function $g_n (x)$ has a slightly more complicated expression
\begin{equation}
    \begin{split}
        g_n \left(x \right) &= \frac{e^{-x}}{\sqrt{x}} \sum_{m=0}^{(n-1)/2} \gamma_m^{\left( n \right)} x^{-m} \\
        &+ \sqrt{\pi} \erf\left(\sqrt{x}\right) \sum_{m=0}^{(n-1)/2} \delta_m^{\left( n \right)} x^{-m},
    \end{split}
\end{equation}
where
\begin{align}
    \gamma_m^{\left( n + 2 \right)} &= \frac{2}{n+1} \left( \delta_m^{\left( n \right)}+  \left(n - m + 1 \right) \gamma_{m-1}^{\left( n \right)}\right), \\
    \delta_m^{\left( n + 2 \right)} &= \frac{2}{n+1} \left( \delta_m^{\left( n \right)}+ \left(n - m + \frac{3}{2} \right)\delta_{m-1}^{\left( n \right)}\right), \\
    \gamma_{0}^{\left( 1 \right)} &= \delta_{0}^{\left( 1 \right)} = 1.
\end{align}
Finally, the function $h_n (x)$ can be expressed in a similar form for even $n$,
\begin{equation}
    \begin{split}
        h_n \left(x \right) &= \frac{e^{-x}}{\sqrt{x}} \sum_{m=0}^{n/2} \epsilon_m^{\left( n \right)} x^{-m} \\
        &+ \sqrt{\pi} \erf\left(\sqrt{x}\right) \sum_{m=0}^{n/2} \zeta_m^{\left( n \right)} x^{-m},
    \end{split}
\end{equation}
where
\begin{align}
    \epsilon_m^{\left( n + 2 \right)} &= \frac{2}{n+2} \left( \zeta_m^{\left( n \right)}+  \left(n - m + 1 \right) \epsilon_{m-1}^{\left( n \right)}\right), \\
    \zeta_m^{\left( n + 2 \right)} &= \frac{2}{n+2} \left( \zeta_m^{\left( n \right)}+ \left(n - m + \frac{3}{2} \right)\zeta_{m-1}^{\left( n \right)}\right), \\
    \epsilon_{0}^{\left( 0 \right)} & = 0, \\
    \zeta_{0}^{\left( 0 \right)} &= \frac{1}{2}.
\end{align}

Whenever $x > 1$, we evaluate the likelihood using \cref{eq:alternative likelihood} and the series expansions for $g_n(x)$ and $h_n(x)$ given above. Otherwise, the likelihood is evaluated using \cref{eq:Hyper likelihood} directly, using the usual series expansion for $_1F_1$
\begin{equation}
    _1F_1 \left( a, c, x \right) = \sum_{m=0}^{\infty} \frac{\left(a \right)_m}{\left(c \right)_m} \frac{x^m}{m!}.
\end{equation}

\end{document}